\documentclass[journal,twoside,web]{ieeecolor}
\newcommand{\normal}{\mathcal{N}}
\newcommand{\complex}{\mathbb{C}}
\newcommand{\Expect}{\mathbb{E}}

\newcommand{\ie}{{\it i.e.}}

\usepackage{tmi}

\usepackage{tikz}

\newcommand\copyrighttext{%
  \footnotesize \textcopyright \the\year{} IEEE. Personal use of this material is permitted. Permission from IEEE must be obtained for all other uses, including reprinting/republishing this material for advertising or promotional purposes, collecting new collected works for resale or redistribution to servers or lists, or reuse of any copyrighted component of this work in other works.
  Citation information: DOI 10.1109/TMI.2024.3443292}

\newcommand\copyrightnotice{%
\begin{tikzpicture}[remember picture,overlay]
\node[anchor=south, yshift=2.5] at (current page.south) {\fbox{\parbox{\dimexpr0.99\textwidth-\fboxsep-\fboxrule\relax}{\copyrighttext}}};
\end{tikzpicture}%
}

\usepackage{etoolbox}
\makeatletter
\@ifundefined{color@begingroup}%
  {\let\color@begingroup\relax
   \let\color@endgroup\relax}{}%
\def\fix@ieeecolor@hbox#1{%
  \hbox{\color@begingroup#1\color@endgroup}}
\patchcmd\@makecaption{\hbox}{\fix@ieeecolor@hbox}{}{\FAILED}
\patchcmd\@makecaption{\hbox}{\fix@ieeecolor@hbox}{}{\FAILED}
\usepackage{cite}
\usepackage{amsmath,amssymb,amsfonts}
\usepackage[ruled,vlined]{algorithm2e}
\usepackage{graphicx}
\usepackage{textcomp}
\usepackage{booktabs}       
\usepackage{multirow}
\usepackage{array}
\usepackage{hyperref}
\def\BibTeX{{\rm B\kern-.05em{\sc i\kern-.025em b}\kern-.08em
    T\kern-.1667em\lower.7ex\hbox{E}\kern-.125emX}}
\begin{document}
\title{AutoSamp: Autoencoding k-space Sampling via Variational Information Maximization for 3D MRI}
\author{Cagan Alkan, Morteza Mardani, \IEEEmembership{Senior Member, IEEE}, Congyu Liao, Zhitao Li, Shreyas S. Vasanawala, and John M. Pauly, \IEEEmembership{Fellow, IEEE}
\thanks{This work was supported by NIH grants R01EB009690, R01EB026136 and U01EB029427.}
\thanks{C. Alkan and J.M. Pauly is with the Department of Electrical Engineering, Stanford University, Stanford, California, USA \mbox{(e-mails:\{calkan,pauly\}@stanford.edu)}.}
\thanks{M. Mardani is with NVIDIA Inc, and the Department of Electrical Engineering, Stanford University, Stanford, California, USA \mbox{(e-mail:mormardani@gmail.com)}.}
\thanks{Z. Li is with the Department of Radiology, Northwestern University, Evanston, Illinois, USA \mbox{(e-mail:zhitaoli@northwestern.edu)}.}
\thanks{C. Liao and S.S. Vasanawala is with the Department of Radiology, Stanford University, Stanford, California, USA \mbox{(e-mails:\{cyliao,vasanawala\}@stanford.edu)}.}
}
\maketitle
\copyrightnotice
\begin{abstract}
Accelerated MRI protocols routinely involve a predefined sampling pattern that undersamples the \mbox{k-space}.
Finding an optimal pattern can enhance the reconstruction quality, however this optimization is a challenging task. To address this challenge, we introduce a novel deep learning framework, AutoSamp, based on variational information maximization that enables joint optimization of sampling pattern and reconstruction of MRI scans. We represent the encoder as a non-uniform Fast Fourier Transform that allows continuous optimization of k-space sample locations on a non-Cartesian plane, and the decoder as a deep reconstruction network. Experiments on public 3D acquired MRI datasets show improved reconstruction quality of the proposed AutoSamp method over the prevailing variable density and variable density Poisson disc sampling for both compressed sensing and deep learning reconstructions. We demonstrate that our data-driven sampling optimization method achieves 4.4dB, 2.0dB, 0.75dB, 0.7dB PSNR improvements over reconstruction with Poisson Disc masks for acceleration factors of \mbox{R = 5, 10, 15, 25}, respectively. Prospectively accelerated acquisitions with 3D FSE sequences using our optimized sampling patterns exhibit improved image quality and sharpness. Furthermore, we analyze the characteristics of the learned sampling patterns with respect to changes in acceleration factor, measurement noise, underlying anatomy, and coil sensitivities. We show that all these factors contribute to the optimization result by affecting the sampling density, k-space coverage and point spread functions of the learned sampling patterns.
\end{abstract}

\begin{IEEEkeywords}
Deep learning, Image reconstruction, \mbox{k-space sampling}, Magnetic resonance imaging, Sampling pattern optimization 
\end{IEEEkeywords}

\section{Introduction}
\label{sec:introduction}
\IEEEPARstart{M}{agnetic} resonance imaging (MRI) reconstruction is an inverse problem where the goal is to reconstruct high fidelity images from spatial Fourier domain (k-space) measurements. MRI scan time is typically reduced by collecting undersampled k-space measurements. The design of the sampling pattern that specifies the set of k-space representations for MRI data acquisition is critical as it directly affects the scan time. Shortening the MRI scan time can have significant benefits such as cutting down MRI scan costs, increasing patient comfort and reducing the effects of motion during scans.

For accelerated MRI, compressed sensing (CS) approaches \cite{lustig2007sparse, lustig2010spirit} rely on incoherence property and leverage pseudo-random sampling patterns. The (incoherent) aliasing artifacts introduced into images via the undersampled measurements are then removed with a non-linear reconstruction algorithm to retrieve high-quality images. Even though the theoretical findings require the restricted isometry property (RIP) and its variants \cite{candes2006stable,donoho2006compressed} to be satisfied for the sampling patterns, CS studies in MRI literature utilize empirical designs such as variable density random patterns \cite{lustig2007sparse}, where the sampling density decays with the distance from k-space origin, and variable density Poisson disc patterns \cite{bridson2007fast, vasanawala2011practical, lustig2010spirit}, which enforce a minimum distance constraint between neighboring samples along with the variable density criterion.

There are a few studies that optimize k-space data sampling patterns for CS reconstruction. \cite{knoll2011adapted} generates sampling patterns using the power spectra of existing reference data sets. Bayesian experimental design framework \cite{seeger2010optimization} uses the information gain criterion to sequentially select the candidate samples for sparse reconstruction under the single-coil setting. They apply their algorithm for Cartesian line selection and Archimedian spiral interleaf angle selection. Methods described in \cite{gozcu2018learning,sanchez2020scalable} are greedy subset selection approaches that choose the sample, line or spoke with the highest performance improvement in the training set at each iteration. \cite{zibetti2021fast} proposes a more efficient algorithm that has faster convergence speed compared with the previous greedy subset selection methods. The authors test their approach by optimizing patterns for different parallel imaging and compressed sensing (PI-CS) and low-rank reconstruction methods. Authors of \cite{haldar2019oedipus} design sampling patterns based on the constrained Cramer-Rao bound. SPARKLING \cite{lazarus2019sparkling} generates trajectories that conform to a heuristically chosen density while achieving a locally uniform k-space coverage that avoids large gaps.

Recently, deep learning (DL) methods have shown promise at solving undersampled MRI reconstruction problems \cite{liang2020deep, hammernik2023physics, knoll2020deep}. DL-based reconstruction algorithms span a wide range of methods that include direct inversion and interpolation via convolutional neural networks (CNN) \cite{wang2016accelerating,hyun2018deep,lee2018deep,han2019k,eo2018kiki,akccakaya2019scan,zhu2018image}, unrolled networks that combine data-consistency blocks with deep-learned regularizers \cite{hammernik2018learning,schlemper2017deep,schlemper2018deep,yang2018admm,zhang2018ista,cheng2019model,aggarwal2018modl,cheng2018highly,sandino2020compressed}, generative adversarial networks (GAN) \cite{yang2017dagan,quan2018compressed,mardani2018deep} and untrained neural networks \cite{darestani2021accelerated,yoo2021time}. Even though these methods have shown improved image quality when solving MR reconstruction problems, sampling patterns for DL reconstruction problems are typically chosen heuristically as in CS studies. The reconstruction models are optimized for a pre-determined acquisition (encoding) model without taking advantage of possible gains that can be obtained via learning the undersampling patterns.

More recently, end-to-end deep learning methods have been proposed for learning undersampling patterns for MRI reconstruction problem. DL-based sample optimization strategies require a fully sampled dataset to simulate the data sampling procedure and evaluate the reconstruction quality. Active acquisition strategies \cite{jin2019self,zhang2019reducing,pineda2020active,bakker2020experimental,van2021active} attempt to predict the next k-space samples to be acquired using information from existing samples. These methods usually employ an additional neural network that select the next sample by measuring the reconstruction quality or estimating uncertainty during the acquisition. In addition, some active acquisition techniques leverage reinforcement learning (RL) based formulations \cite{pineda2020active,bakker2020experimental,van2021active}. 

The non-active strategies can be grouped into two categories. The probabilistic approaches \cite{bahadir2019learning, zhang2020extending, huijben2020learning} focus on the Cartesian sampling case and model binary sampling masks probabilistically. These methods employ certain relaxations such as straight-through estimator \cite{bengio2013estimating} and Gumbell-Softmax re-parameterization \cite{jang2016categorical} to maintain differentiability with respect to sampling probabilities. They generally treat the sampling points independently and do not model the dependencies between sampling locations explicitly. The second type of methods \cite{aggarwal2020j,weiss2019pilot} directly optimize for the k-space coordinates instead of estimating sampling probabilities. Authors of J-MoDL \cite{aggarwal2020j} restrict the optimization to a set of separable variables such as horizontal and vertical directions in 2-D plane to reduce the search space. As a result, their optimized sampling masks are restricted to a uniform {\it grid} pattern and do not cover the more general sampling patterns. A few other works incorporate MR gradient system amplitude and slew rate constraints into their loss functions to optimize sampling {\it trajectories}. \cite{weiss2019pilot} proposes an end-to-end multi-scale and multi-stage approach that optimizes the sampling trajectories for reconstruction and segmentation tasks. \cite{wang2022b} parameterizes the trajectories with quadratic B-spline kernels for 2D MRI and uses multi-level optimization to both stabilize the training process and satisfy the constraints more easily. \cite{wang2022stochastic} further extends this work for 3D non-Cartesian sampling trajectories. The authors also model peripheral nerve stimulation (PNS) and include it as a soft constraint in the loss function to suppress its effect. \cite{peng2022learning} uses a neural Ordinary Differential Equation (ODE) solver to approximate the trajectory dynamics and optimizes its parameters along with a DL reconstruction network.

In this work, we present a variational information maximization method that enables joint optimization of acquisition and reconstruction of MRI scans in an end-to-end, data-driven manner. Our method enables learning an undersampling pattern tuned specifically to the reconstruction network, and vice versa, to obtain improved reconstruction performance. We consider the optimization of phase encoding coordinates in 2D k-space, which corresponds to the 3D imaging scenario with a fully sampled readout axis. We represent the acquisition (encoder) model with the non-uniform Fast Fourier Transform (nuFFT) operator \cite{beatty2005rapid,greengard2004accelerating} that is parameterized by the sampling locations in k-space. This allows interpolation of non-Cartesian coordinates and enables continuous optimization of the sampling pattern. On the reconstruction (decoder) side, we use an unrolled reconstruction network which mimics the proximal gradient based solutions to compressed sensing problems. We then analyze the learned sampling patterns with respect to changes in acceleration factor, measurement noise, dataset and coil sensitivities. Our contributions can be summarized as follows:
\begin{itemize}
    \item We propose a novel and versatile framework based on variational information maximization that facilitates joint optimization of MRI data sampling and reconstruction.
    \item We demonstrate that our data-driven sampling optimization method shows $4.4$dB, $2.0$dB, $0.75$dB, $0.7$dB PSNR improvement over reconstruction with Poisson Disc masks for $R=5,10,15,25$, respectively.
    \item We implement our optimized sampling patterns with a 3D fast spin echo (FSE) sequence and illustrate the improved image quality and sharpness on prospectively accelerated acquisitions.
    \item We conduct empirical analysis to investigate how the learned sampling patterns depend on various factors, including acceleration factor, measurement noise, dataset, and coil sensitivities.
\end{itemize}

\section{Methods}
\subsection{Problem Setting}
\begin{figure}[h!]
\centerline{\includegraphics[width=1.05\columnwidth]{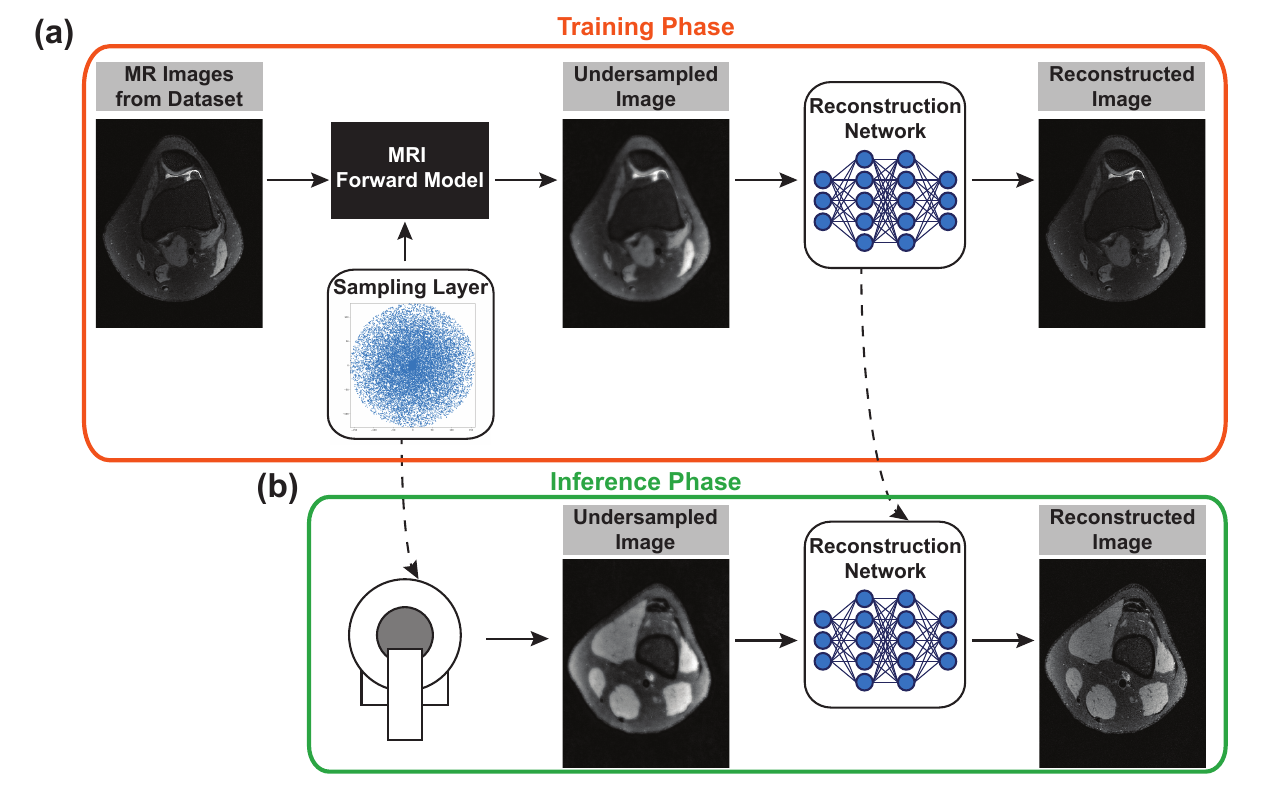}}
\caption{Training and inference pipeline for AutoSamp: In the training phase (a), AutoSamp jointly learns a sampling pattern and a reconstruction network using a fully-sampled dataset. During the inference phase (b), the learned pattern and reconstruction network are fixed and used to collect and reconstruct new data.}
\label{fig:end_to_end}
\end{figure}

We consider the MR signal model under the additive white complex Gaussian noise as 
\begin{align}\label{eq:1}
    z = f_\phi(x) + \epsilon
\end{align}
where $x\in\complex^N$ is the image, $z\in\complex^M$ is the measured data in k-space domain, $f_\phi(\cdot)$ is the forward model that describes the imaging system parameterized by k-space sample coordinates $\phi \in [-0.5,0.5]^M$, and $\epsilon \sim \normal_c(0,\sigma^2I)$ is the measurement noise. The imaging model $f_\phi$ includes non-uniform Fast Fourier Transform (nuFFT) operation $F_{nu}$, and for the multi-coil scenario, contains signal modulation by coil sensitivity maps $S$. Given an acceleration factor $R=N/M$, our goal is to find the optimal set of samples $\phi$ along with a reconstruction function $g(z)$ that maintains the full k-space data image quality. As shown in Fig. \ref{fig:end_to_end}, we jointly learn the optimal sampling pattern and reconstruction function using a dataset during the training phase. We then fix this learned pattern to collect data from the MR scanner and use the reconstruction network to reconstruct new data.

\subsection{Variational Information Maximization for Acquisition and Reconstruction}
Under the setting in \eqref{eq:1}, the distribution of measurements $Z$ given image $X$ has a complex valued Gaussian distribution $q_\phi(Z|X)=\normal_c(f_\phi(X), \sigma^2I)$. Here, the parameters $\phi$ correspond to the coordinates of the points in the sampling pattern.
Our goal is to learn the parameters $\phi$ that allow high fidelity recovery of image $X$ given measurements $Z$. We adapt the uncertainty autoencoder framework defined in \cite{grover2019uncertainty} and we make use of the InfoMax principle \cite{linsker1988self, linsker1989generate} that maximizes the mutual information between the measurements and noisy latent representations
\begin{align}
    \max_\phi I_\phi(X,Z) &= \max_\phi \int q_\phi (x,z) \log{\frac{q_\phi(x,z)}{q(x)q_\phi(z)}} \,dx\,dz \\
    &= \max_\phi H(X)-H_\phi(X|Z)
\end{align}
where $I_\phi(X,Z)$ denotes the mutual information between $X$ and $Z$, and $H$ represents differential entropy. This formulation enforces the measurement space $Z$ to have maximal information about the signal $X$ so that the image reconstruction has a smaller error.
Since the data entropy $H(X)$ does not depend on k-space coordinates $\phi$, this is equivalent to
\begin{align}
    \max_\phi -H_{\phi}(X|Z) &= \max_\phi \Expect_{q_{\phi}(X,Z)}[\log q_\phi(X|Z)] \label{eq:3}\\
    &\geq \max_{\phi,\theta} \Expect_{q_{\phi}(X,Z)}[\log p_\theta(X|Z)] \label{eq:4}
\end{align}
 The conditional entropy in \eqref{eq:3} requires evaluating the model posterior $q_\phi(X|Z)$, however this term is intractable due to the high dimensionality of $X$ and $Z$. Therefore, in  \eqref{eq:4} we introduce a variational approximation $p_\theta(X|Z)$ to the posterior distribution via a new set of parameters ($\theta$) and obtain a lower bound \cite{barber2004algorithm}. The variational parameters ($\theta$) correspond to the weights of the decoder, i.e., the reconstruction network. Using the bound on \eqref{eq:4}, we express our objective as
\begin{align}
    \max_{\phi, \theta} \mathcal{L}(\phi, \theta; \mathcal{D}) = \max_{\phi, \theta} \sum_{x\in\mathcal{D}} \Expect_{q_\phi(Z|x)}[\log p_\theta(x|z)] \label{eq:5}
\end{align}
where we estimate the expectation with respect to $q_\phi(X)$ using Monte-Carlo sampling from the dataset $\mathcal{D}$. Thus, we propose optimizing the shared $\phi$ and $\theta$ for the entire dataset rather than specific acquisition and reconstruction parameters for each example in the dataset. In addition, for a different anatomy (hence for a different dataset), the optimization must be repeated.

Depending on the observation model used for $p_\theta(\cdot)$, we end up with a different loss function. For example, in the Gaussian assumption case $p_\theta(X|z)=\normal_c(p_\theta(z), \tilde{\sigma}^2 I)$, we have
\begin{align}
    \mathcal{L}(\phi,\theta,\mathcal{D}) = \sum_{x\in\mathcal{D}} \Expect_{q_\phi(Z|x)}[\| x-p_\theta(z)\|_2^2].
\end{align}
Similarly, for the Laplacian observation model, the loss function is the $\ell_1$ distance between reconstruction and ground truth.

We compute the expectation with respect to $q_\phi(Z|x)$ using Monte-Carlo methods, and estimate the gradients with respect to $\phi$ using reparameterization trick \cite{kingma2013auto} which allows us to calculate this gradient as
\begin{align}
    \nabla_{\phi} \Expect_{q_\phi(Z|x)}[\log p_\theta(x|z)] &= \nabla_{\phi} \Expect_{p(\epsilon)} [\log p_\theta(x|z = f_{\phi}(x) + \epsilon)]\\
    &= \Expect_{p(\epsilon)} [\nabla_{\phi}\log p_\theta(x|z = f_{\phi}(x) + \epsilon)]
\end{align}

The main difference between \cite{grover2019uncertainty} and our formulation is that we enforce the latent space to correspond to the Fourier domain coefficients by using a nuFFT operator rather than a fully parameterized neural network. This allows us to optimize the sampling pattern and reconstruction network in alignment with the MRI signal model.
\subsection{Multi-Channel Acquisition and Reconstruction Models}
\begin{figure*}[h!]
\centerline{\includegraphics[width=0.8\linewidth]{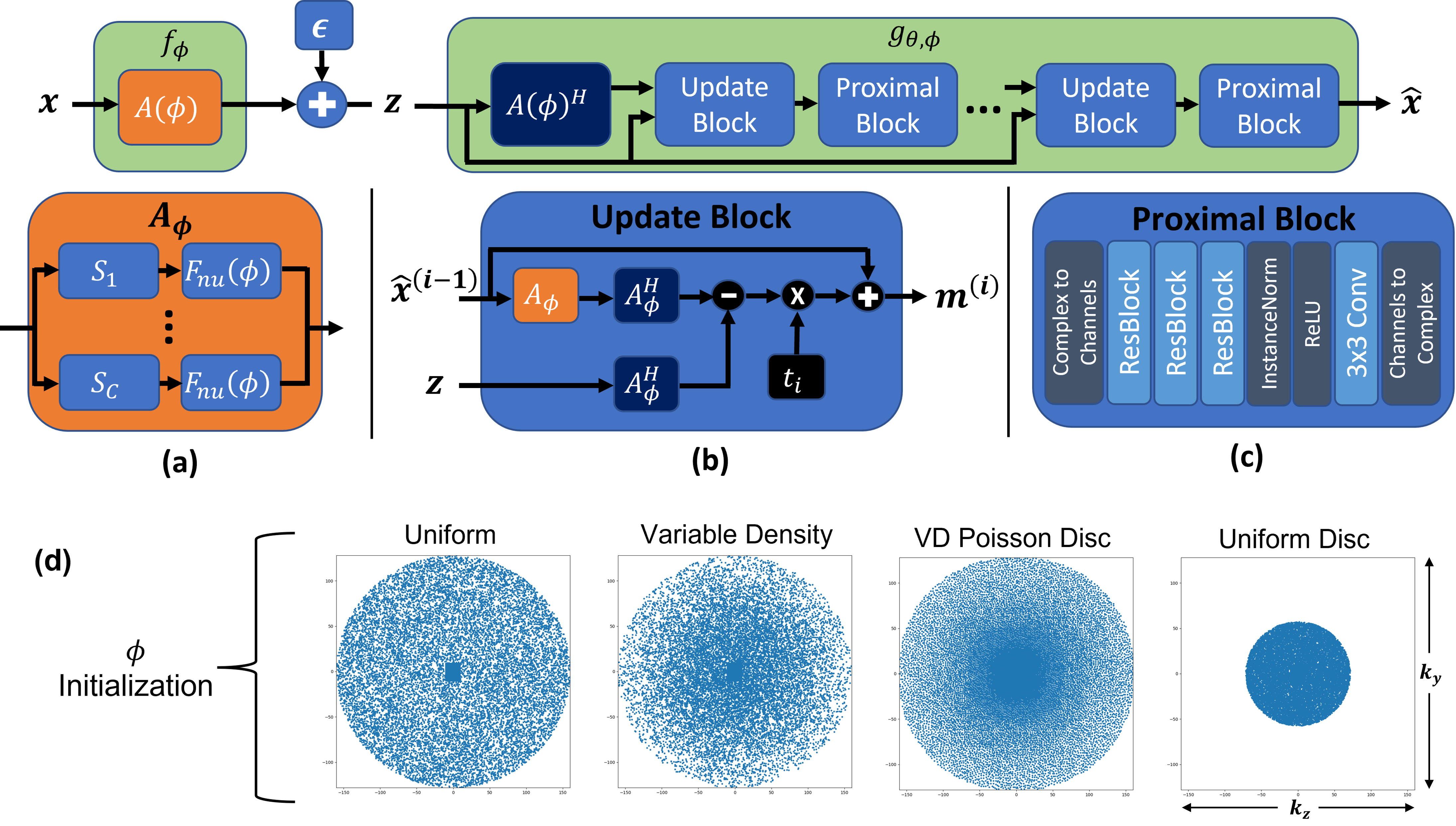}}
\caption{Network architecture consists of a nuFFT based encoder (a) and an unrolled reconstruction network (b, c). $x$ is the image from the dataset, $z$ is the measured data in k-space domain and $\hat{x}$ is the reconstruction. Update block (b) consists of data-consistency steps using the measurements $z$, and the proximal block (c) consists of multiple residual blocks. The sampling coordinates ($\phi$) are shared between encoder and decoder, and initialized with one of the patterns shown in (d).}
\label{fig:architecture}
\end{figure*}

Our overall network architecture is illustrated in Figure \ref{fig:architecture}. We represent the acquisition (encoding) model with the nuFFT operator that is parameterized by the sampling locations $\phi$ in k-space. Using nuFFT instead of an FFT enables us to represent the k-space sampling locations as continuous variables. As a result, we directly optimize the sampling pattern by backpropagating through $\phi$. In the multi-channel MRI setting, the acquisition model admits
\begin{align}\label{eq:6}
    f_\phi(x) = A_\phi x= \left[\begin{array}{ccc}
         F_{nu}(\phi)S_1 \\
         \vdots \\
         F_{nu}(\phi)S_C
    \end{array}\right] x
\end{align}
where $C$ is the number of channels, $S_i \in \complex^{N\times N}$ is a diagonal matrix containing coil sensitivity profiles for coil $i$, and $F_{nu}(\phi): \complex^N\rightarrow\complex^{M/C}$ is the nuFFT operator at sampling locations specified by $\phi$.

Notice that \eqref{eq:5} allows using any deep neural network for reconstruction as the decoder. In this work, we used an unrolled reconstruction network (UNN) similar to \cite{cheng2018highly,sandino2020compressed,mardani2018neural} that mimic the proximal gradient descent (PGD) based solutions for non-smooth compressed sensing. The unrolled architecture also incorporates MR physics, where the reconstruction is performed by alternating between data consistency (Fig. \ref{fig:architecture}b) and proximal steps (Fig. \ref{fig:architecture}c) for a fixed number of iterations. More specifically, the unrolled PGD expresses the output $\hat{x}^{(i)}$ at the $i$th iteration as:
\begin{align}
    \hat{x}^{(i)} &= g_{\theta_i}\left(\mathbf{DC}(\hat{x}^{(i-1)}, z; \phi)\right)\\
              &= g_{\theta_i}\Big(\underbrace{\hat{x}^{(i-1)}-2tA_{\phi}^H(A_{\phi}\hat{x}^{(i-1)}-z)}_{m^{(i)}}\Big)
\end{align}
where $g_{\theta_i}$ is the neural network and $\theta_i$ are its parameters at the $i$th iteration for $i\in\{1,\ldots,n\}$. The zero-filled reconstruction $\hat{x}^{(0)} = A_\phi^Hz$ is used as the input to the first data consistency block.
The combination of intermediate data consistency steps and the proximal blocks renders the reconstruction a function of both $\phi$ and $\theta$, hence we share the parameters of the nuFFT encoder with the decoder network. The unrolled nature of the network also allows for multiple gradient paths to $\phi$ from the decoder. Algorithm \ref{alg:autosamp} describes the learning process for AutoSamp.
\RestyleAlgo{ruled}
\SetKwComment{Comment}{/* }{ */}
\SetKwInOut{KwInit}{Initializate}
\SetKw{KwRet}{Return: }
\begin{algorithm}
\caption{Training process for AutoSamp}\label{alg:autosamp}
\KwIn{Fully-sampled dataset $\mathcal{D}$; MRI forward model $f_\phi(\cdot)$ using nuFFT; Observation model $p_\theta(\cdot)$ that includes the reconstruction network $g_\theta$; Measurement noise level $\sigma$; Number of epochs $N_{\texttt{epoch}}$; Learning rates  $\eta_\phi$, $\eta_\theta$ }
\KwInit{$\phi^{0}$, $\theta^{0}$}
\For{$i=0$ to $N_{\texttt{epoch}}-1$}{
    \For{$x$ in $\mathcal{D}$}{
        // Simulate measurement noise\\
        $\epsilon \sim \normal_c(0,\sigma^2I)$ \\
        // Apply forward model and add noise\\
        $z = f_{\phi}(x) + \epsilon$ \\
        // Reconstruct and calculate loss
        $\mathcal{L} = -\log(p_\theta(x|z))$ \\
        // Update pattern and reconstruction network\\
        $\phi^{i+1} = \phi^{i} - \eta_\phi \nabla_\phi\mathcal{L}$ \\
        $\theta^{i+1} = \theta^{i} - \eta_\theta \nabla_\theta\mathcal{L}$
    }
}
\KwRet{$\phi^{N_{\texttt{epoch}}}$, $\theta^{N_{\texttt{epoch}}}$}
\end{algorithm}

At inference time, the  undersampled k-space data $z$ is directly inputted to the neural network $g_\theta(z)$ to perform reconstruction.

\begin{figure*}[h!]
\centerline{\includegraphics[width=\linewidth]{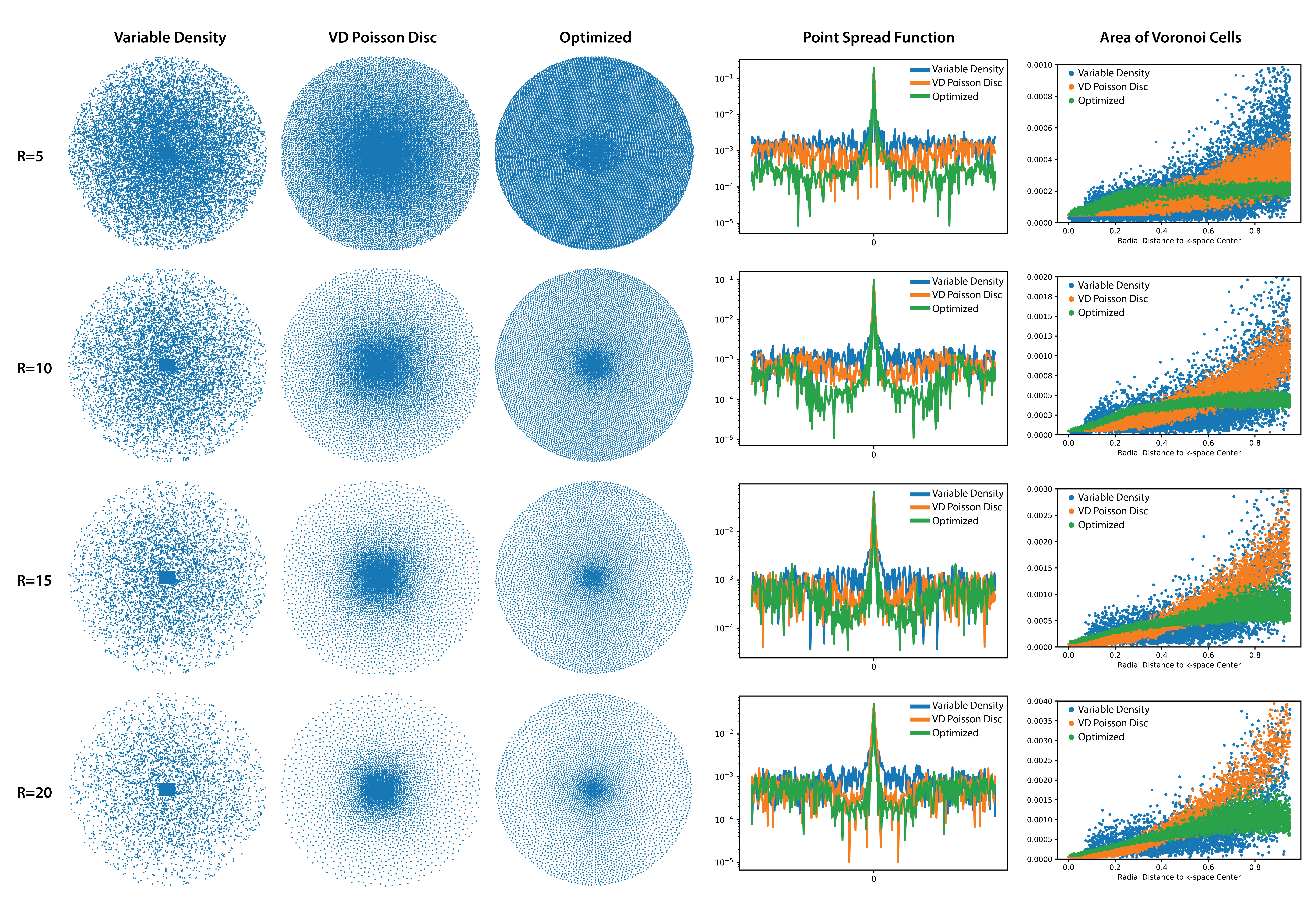}}
\caption{Sampling pattern dependence on acceleration factor and analysis of optimized (learned) sampling patterns. Power spread function (PSF) plots show the profile across a horizontal line through the origin. Learned patterns avoid clustered samples and the gaps between samples are smaller. PSF sidelobes of the optimized patterns are reduced compared to VD and VD Poisson Disc patterns. Sidelobe reduction is much higher for lower acceleration factors. Voronoi cell area distributions indicate learned patterns have more uniform density outside the densely sampled central area. For detailed views of patterns, please refer to the zoom-in views provided in Fig. \ref{fig:zoomin} in Appendix.}
\label{fig:acceleration}
\end{figure*}

\section{Experiments}
In this work, we focus on the optimization of sampling patterns for 3D imaging with Cartesian readouts. We utilized the separable nature of the problem by first taking a 1D Fourier Transform along the readout dimension ($k_x$) to obtain a 3D image in the hybrid $(x,k_y,k_z)$ space. We then treat each \mbox{$x$-slice}, or \mbox{$x$-section}, consisting of $(k_y,k_z)$ dimensions as a separate example.
\subsection{Datasets}
We relied on two datasets for comparing the results.
\subsubsection{Stanford fully-sampled 3D FSE knees dataset}
As our primary dataset, we used the "Stanford Fully Sampled 3D FSE Knees" dataset available in mridata.org \cite{ong2018open} which contains 3D knee scans from 19 subjects. Each 3D volume has 8 channels and consists of 320 data points along the readout dimension $k_x$. The data was acquired in corner-cut fashion in k-space, \ie, the samples falling outside of an ellipse centered at k-space origin are not collected. Each of the $x$-slices were treated as separate examples during training and validation with a matrix size of $256\times320$. The sensitivity maps of each coil were estimated using ESPIRiT \cite{uecker2014espirit} and JSENSE \cite{ying2007joint} algorithms. The datasets were divided according to subjects where 14 subjects (4480 slices) were used for training, 2 subjects (640 slices) were used for validation, and 3 subjects (960 slices) were used for testing.
\subsubsection{Multichannel brain dataset}
We also used the 12-coil 3D T2 CUBE Brain dataset available in \cite{aggarwal2018modl,aggarwal2020j} where 4 subjects are used for training and 1 subject is used for testing. Similar to the knee dataset, the slice axis is chosen to be the readout direction $k_x$, and each $x$-slice was treated as separate example during training and testing. The dataset consists of a total 360 $x$-slices, each one having a shape ($k_y\times k_z$) = ($256\times232$). The coil sensitivities for the brain dataset were estimated using ESPIRiT.
\subsection{Comparison of Learned Sampling Patterns} \label{sec:pattern_comparison}
We compared the patterns learned using our proposed method with Variable Density (VD) \cite{lustig2007sparse}, and VD Poisson Disc \cite{bridson2007fast,vasanawala2011practical,lustig2010spirit} sampling patterns that are commonly used in CS and DL based reconstructions. Both patterns have calibration regions of size $20\times20$ for all acceleration factors. The VD sampling patterns are realized by sampling from a 2D truncated Gaussian distribution. For the VD patterns of the knee dataset, samples that fall outside of the ellipse centered at the k-space origin are rejected and resampled to have the effect of corner cutting. VD Poisson Disc sampling patterns are realized using SigPy \cite{ong2019sigpy}, where the sampling mask has a k-space density proportional to $1/(1+s|r|)$, with $r$ representing the k-space radius, and $s$ representing the slope. The slope is iteratively adjusted until the target acceleration rate is achieved.

Pattern characteristics were analyzed using point spread functions (PSF) \cite{baron2018rapid} that characterize the extent of blurring and aliasing effects and radial distribution of Voronoi cell areas \cite{rasche1999resampling} which is a measure for the sampling density in k-space.

For each of the baseline sampling patterns, we trained unrolled reconstruction networks to compare reconstruction performances. In addition, we implemented the PGD variant of J-MoDL \cite{aggarwal2020j}, which we label as J-PGD, and reported its performance as a DL-based sampling optimization baseline. 
\subsection{Initialization of Sampling Locations}
We considered 4 different initializations for the sampling patterns that are shown in Fig \ref{fig:architecture}d. All of the initializations have $20\times20$ calibration region in k-space.
\begin{itemize}
    \item Uniform initialization: Samples are drawn from a uniform random distribution  defined on a disc (knee dataset) or full k-space extent (brain dataset).
    \item Variable density initialization: VD sampling patterns are generated in the way described in Sec. \ref{sec:pattern_comparison}.
    \item VD Poisson Disc initialization: VD Poisson Disc sampling patterns are generated in the way described in \mbox{Sec. \ref{sec:pattern_comparison}}.
    \item Uniform small disc initialization: Samples are drawn from a uniform random distribution  defined on a disc with half the radius of uniform initialization (knee dataset) or half k-space extent (brain dataset). This initialization provides faster convergence especially for the high noise experiments shown in Sec. \ref{sec:impact_of_noise}.
\end{itemize}
\subsection{Implementation Details}
\subsubsection{nuFFT Library}
We used our own TensorFlow2 implementation of nuFFT library (tfnufft\footnote{https://github.com/alkanc/tfnufft}) that we made publicly available in our codebase. For all of the experiments, our nuFFT implementation uses a Kaiser-Bessel kernel with an oversampling factor of $1.25$ and a width of $4$ in oversampled grid units. Note that our implementation calculates the Kaiser-Bessel kernel values on the fly, and no table interpolation is used. This approach avoids the gradient instabilities mentioned in \cite{wang2023efficient}, and enables the use of automatic differentiation tools for calculating gradients with respect to k-space coordinates.
\subsubsection{Unrolled PGD Model}
The proximal blocks consist of $3$ residual blocks, and we unrolled the network for $n=8$ iterations. The convolutional blocks in each residual block consists of convolutional layers with $128$ channels and a kernel size of $3$ followed by instance normalization and ReLU activation. There is no weight sharing between proximal blocks. The hyperparameters were carefully tuned according to validation set performance within the GPU memory limitations. Complex valued data were represented as separate real and imaginary channels inside the proximal blocks as shown in Fig. \ref{fig:architecture}c.
\subsubsection{Compressed Sensing (CS) Reconstruction}
CS reconstruction was performed with $\ell_1$-wavelet regularization \cite{lustig2007sparse} for 100 iterations for each slice using SigPy \cite{ong2019sigpy}. The regularization strength was tuned for different datasets and sampling patterns.
\subsubsection{Training Details}
All of our models were implemented on TensorFlow2 and trained on a NVIDIA GPUs with either 12, 16 or 24 GB of memory. As the sampling locations change during training, we did not use density compensation while calculating the adjoint nuFFT unlike as in \cite{ramzi2021density, ramzi2022nc}. Instead, we relied on the proximal block to compensate for the density.

We considered the Laplacian observation model for $p_\theta(\cdot)$ in Equation \ref{eq:5} which corresponds to $\ell_1$ reconstruction loss. Due to GPU memory constraints, we used a batch size of 1. We also used the Adam optimizer with $\beta_1 = 0.9$, $\beta_2 = 0.999$. We set the learning rates as $0.01$ and $0.001$ for the sampling points $\phi$ and reconstruction network parameters $\theta$, respectively.

We jointly trained the network and the sampling patterns for 300 epochs ($\approx6$ days) on the knee dataset and for 1250 epochs ($\approx3$ days) on the brain dataset. Once trained, each x-slice takes $\approx430$ms to reconstruct and the reconstruction can be parallelized by batching.

To assess the image quality, we adopt peak signal to noise ratio (PSNR) and structural similarity index measure (SSIM) between the reconstruction and fully-sampled ground truth. PSNR was evaluated directly on complex-valued images, whereas SSIM was calculated on magnitude images. Reported metrics were computed for each slice instead of the entire volume. At inference time, the model checkpoint that achieves the highest validation PSNR metric was selected.

Our implementation is publicly available at \url{https://github.com/alkanc/autosamp}.

\subsection{Prospective Studies}
We implemented the optimized sampling patterns prospectively with a proton density weighted fat-saturated 3D fast spin echo (FSE) CUBE \cite{busse2008effects} sequence on a 3T MRI system (GE Healthcare) equipped with an 18-channel knee coil. To compare the image quality, we also implemented prospective VD Poisson disc sampling patterns with the same sequence. In order to match the parameters of the Stanford 3D FSE knee dataset, scan parameters were chosen as TR/TE=1550/25ms, ETL=40, BW= 50kHZ with a FOV of 160mm x 160mm and a slice thickness of 0.6mm. The echoes in the echo train are ordered for radial modulation to achieve proton density-weighted contrast \cite{busse2008effects}. Additionally, we acquired a low-resolution Cartesian GRE image in order to generate coil sensitivity maps. The sensitivity maps were estimated using ESPIRiT \cite{uecker2014espirit} algorithm and the same set of sensitivity maps were used in all reconstructions. All imaging was performed with Stanford Institutional Review Board (IRB) approval (protocol \#63570) and consent.

We acquired data for $R=10$ and $R=20$ acceleration factors with scan times of 05:18 and 02:40, respectively. Corresponding fully-sampled $R=1$ acquisition would take 53:20. The acquired data were reconstructed with both UNN and CS methods. We then compared the image quality between the images acquired with AutoSamp-optimized and VD Poisson Disc patterns.

\section{Results}

\subsection{Reconstruction Quality Comparison} \label{sec:recon_quality}

\begin{figure*}[h!]
\centerline{\includegraphics[width=0.99\linewidth]{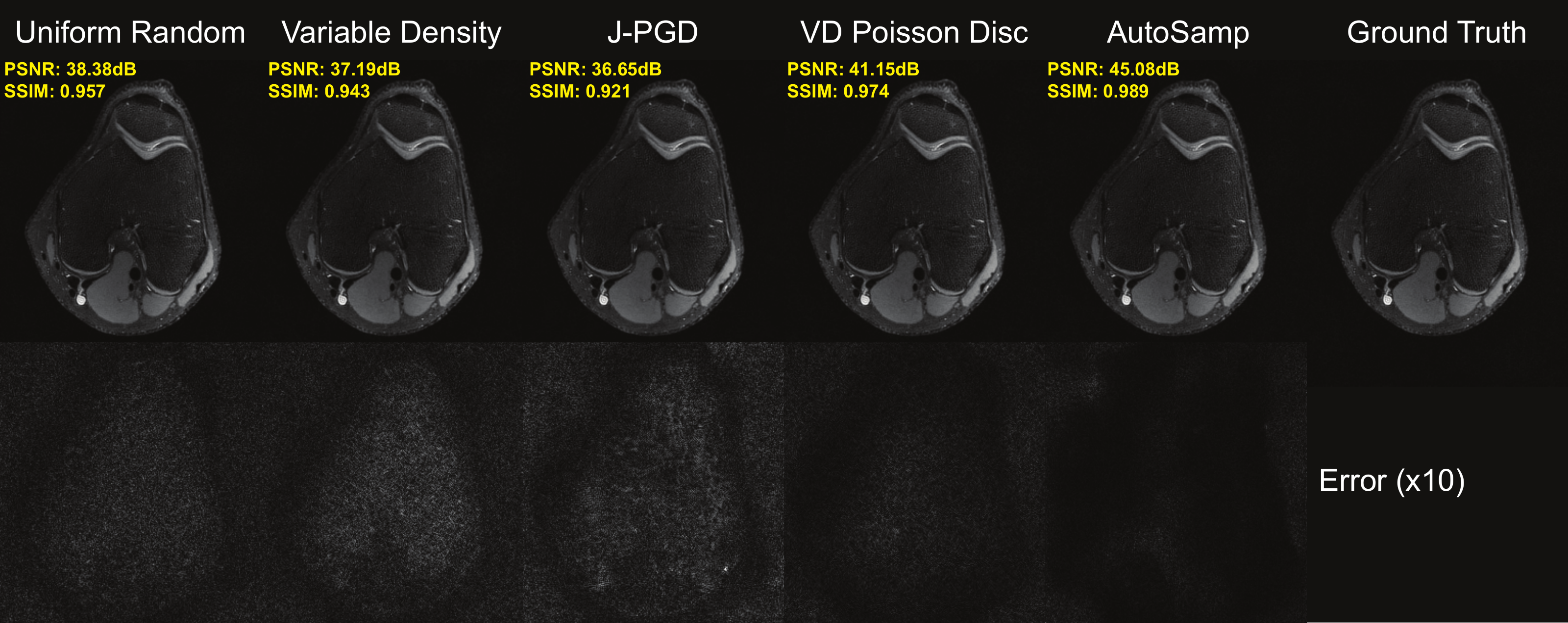}}
\caption{Reconstructed images from an example slice from the test set for $R=5$. Error maps indicate that sample optimization using our proposed method (AutoSamp) has the highest reconstruction quality.}
\label{fig:reconimages_5x}
\end{figure*}

\begin{figure*}[h!]
\centerline{\includegraphics[width=0.99\linewidth]{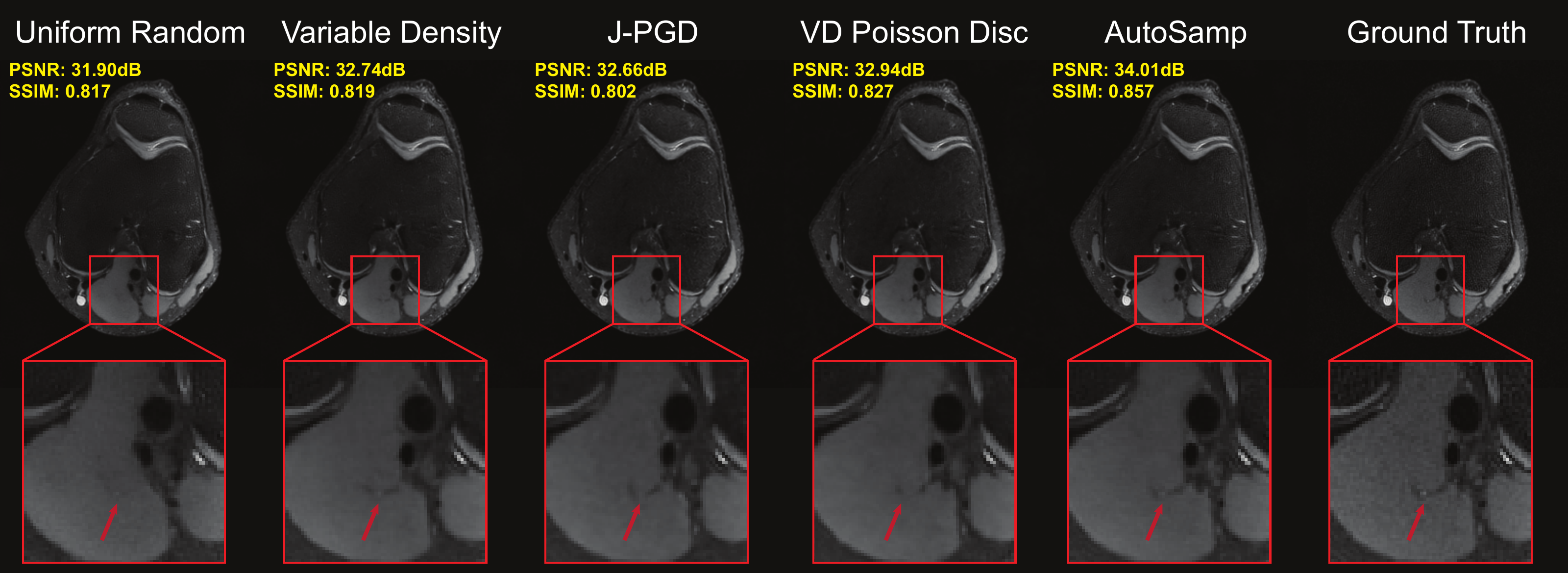}}
\caption{Reconstructed images from an example slice from the test set for $R=20$. Zoomed-in regions show that sample optimization using our proposed method (AutoSamp) preserves structural details better and improves overall visual quality compared to other methods.}
\label{fig:reconimages_20x}
\end{figure*}

We first run experiments for four acceleration factors \mbox{$R=\{5,10,15,20\}$} with the noise variance $\sigma=0.0001$ in Eq. \ref{eq:1}. Table \ref{table:metrics} reports the average quantitative metrics on the Stanford Knee Test Dataset. Note that the patterns from \mbox{J-PGD} and our proposed AutoSamp method were optimized jointly with the UNN reconstruction, whereas only the UNN was optimized for the VD and VD Poisson Disc patterns. All of the metric values for AutoSamp in this table are obtained with uniform initialization as it provided the highest validation metrics. The results from UNN reconstructions indicate that the proposed method improves both PSNR and SSIM values for all acceleration factors compared to other methods. Metric improvements are much higher for lower acceleration factors. Optimized patterns show $4.4$dB, $2.0$dB, $0.75$dB, $0.7$dB PSNR improvement over reconstruction with Poisson Disc masks in the test set for $R=5,10,15,25$, respectively. Reconstructed images from an example slice from the test set are shown in Fig. \ref{fig:reconimages_5x} and \ref{fig:reconimages_20x} for acceleration factors $R=\{5,20\}$. The error maps indicate that our proposed method reduces reconstruction errors, improving overall quality. In addition, zoomed-in regions in Fig. \ref{fig:reconimages_20x} illustrate that our method preserves fine structural details better than the other methods.

We then assess the generalizability of the patterns by testing their performance for CS reconstruction. The bottom portion of Table \ref{table:metrics} reports CS reconstruction metrics on the test set. Even though the patterns from our proposed method were optimized for PGD based UNN reconstruction, they still improve the reconstruction quality for CS reconstruction.

Appendix~\ref{appendix:opt_strategy} shows the benefits of simultaneously optimizing sampling and reconstruction for UNN reconstructions.

\begin{table}[h!]
\caption{Average quantitative reconstruction quality on the Stanford Knee test set}
\label{table:metrics}
\centering
\resizebox{\columnwidth}{!}{%
\begin{tabular}{cllll}
\toprule
Reconstruction & Acceleration ($R$) & Sampling Pattern & PSNR & SSIM \\
\midrule
\multirow{16}{*}{UNN} 
    & \multirow{4}{*}{5x} & Variable Density & 36.43 & 0.928\\
        & & VD Poisson Disc & 39.75 & 0.964\\
        & & J-PGD & 35.69 & 0.905\\
        & & AutoSamp (UNN) & \bfseries44.24 & \bfseries0.988\\\cmidrule{2-5}
    & \multirow{4}{*}{10x} & Variable Density & 33.82 & 0.862\\
        & & VD Poisson Disc & 33.94 & 0.872\\
        & & J-PGD & 32.95 & 0.825\\
        & & AutoSamp (UNN) & \bfseries35.95 & \bfseries0.909\\\cmidrule{2-5}
    & \multirow{4}{*}{15x} & Variable Density & 32.72 & 0.817\\
        & & VD Poisson Disc & 33.13 & 0.829\\
        & & J-PGD & 31.90 & 0.767\\
        & & AutoSamp (UNN) & \bfseries33.87 & \bfseries0.859\\\cmidrule{2-5}
    & \multirow{4}{*}{20x} & Variable Density & 31.97 & 0.781\\
        & & VD Poisson Disc & 32.33 & 0.790\\
        & & J-PGD & 32.20 & 0.779\\
        & & AutoSamp (UNN) & \bfseries33.03 & \bfseries0.817\\\midrule
\multirow{16}{*}{CS} 
    & \multirow{4}{*}{5x} & Variable Density & 38.12 & 0.955\\
        & & VD Poisson Disc & 40.86 & 0.975\\
        & & J-PGD & 40.78 & 0.974\\
        & & AutoSamp (UNN) & \bfseries45.83 & \bfseries0.993\\\cmidrule{2-5}
    & \multirow{4}{*}{10x} & Variable Density & 33.92 & 0.895\\
        & & VD Poisson Disc & 34.24 & 0.895\\
        & & J-PGD & 34.61 & 0.905\\
        & & AutoSamp (UNN) & \bfseries35.87 & \bfseries0.927\\\cmidrule{2-5}
    & \multirow{4}{*}{15x} & Variable Density & 31.77 & 0.842\\
        & & VD Poisson Disc & 32.29 & 0.845\\
        & & J-PGD & 32.50 & 0.852\\
        & & AutoSamp (UNN) & \bfseries32.93 & \bfseries0.868\\\cmidrule{2-5}
    & \multirow{4}{*}{20x} & Variable Density & 30.41 & 0.796\\
        & & VD Poisson Disc & 31.34 & 0.814\\
        & & J-PGD & \bfseries32.16 & \bfseries0.839\\
        & & AutoSamp (UNN) & 31.63 & 0.831\\\midrule
\end{tabular}
}
\end{table}

\subsection{Interpretation of Optimized Sampling Patterns}

Fig. \ref{fig:acceleration} displays the sampling pattern comparisons for \mbox{$R=\{5,10,15,20\}$} along with the point spread functions and Voronoi cell area distributions. We observe that the optimized patterns have dense sampling around k-space origin. The optimized samples avoid clustering, but at the same time the large gaps between neighboring samples are reduced, which is also a feature of Poisson Disc sampling. However, unlike the typical VD and VD Poisson Disc patterns, the sampling density for the optimized patterns decays much more slowly as the distance to k-space origin increases. The Voronoi cell area distributions conform to a $k_r^{1/2}$ relation, where $k_r$ is the k-space radius, especially for $R=5$ and $R=10$. The arrangement of optimized samples resemble hexagonal packing which is known to have the highest-density lattice packing of circles \cite{sloane1984packing}. Point spread function profiles indicate that the sidelobes are suppressed in optimized patterns, which makes the job of the reconstruction network easier and consequently improves the reconstruction performance. In accordance with the gains reported in Table \ref{table:metrics} and Sec. \ref{sec:recon_quality}, sidelobe reduction is much higher for lower acceleration factors. Overall, patterns learned using our proposed algorithm adapts the sample distributions and spacing to the acceleration factor.

\subsection{Dependence on Dataset and Anatomy}
\begin{figure}[h!]
\centerline{\includegraphics[width=1.05\columnwidth]{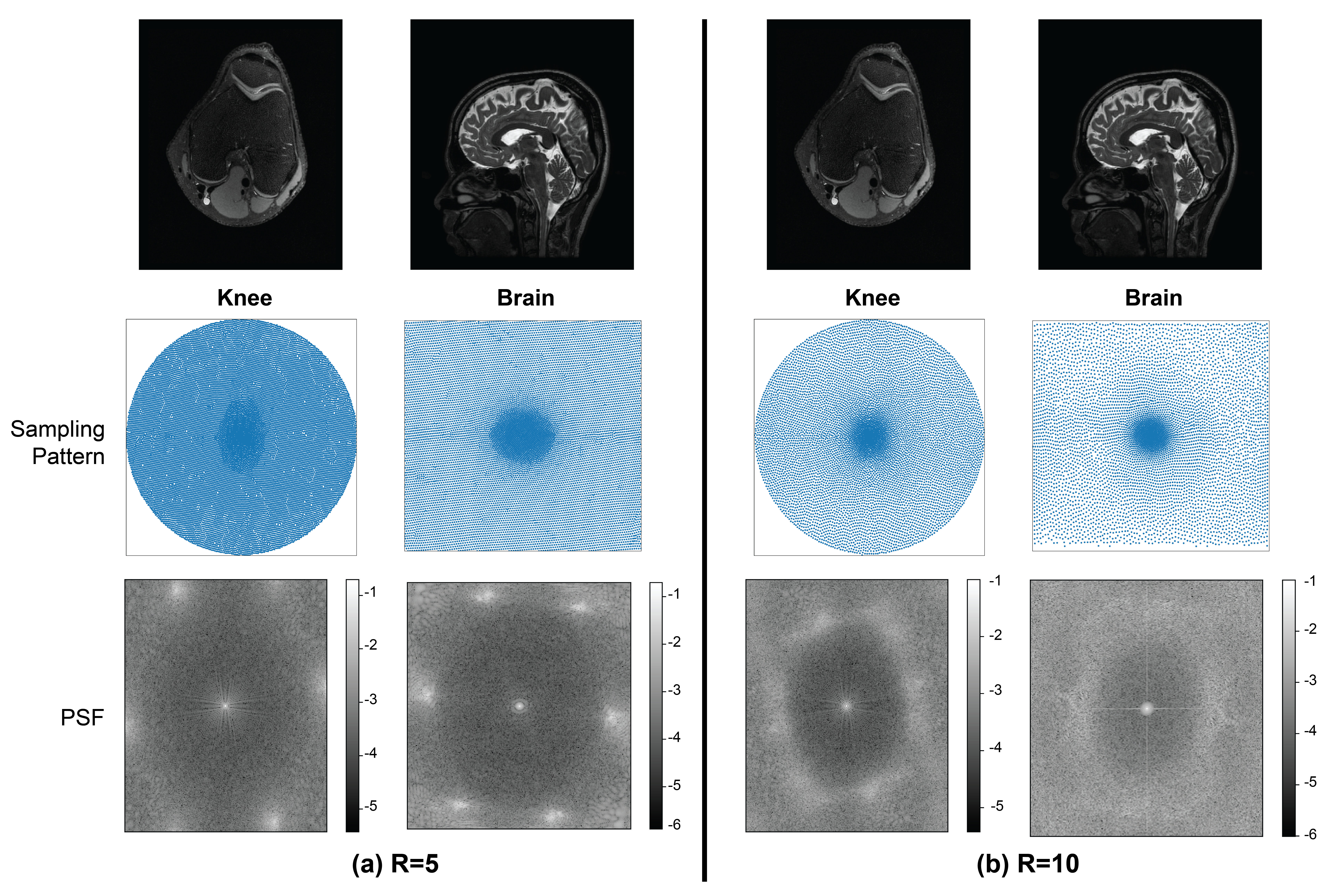}}
\caption{Effect of the dataset (and also the anatomy of interest) on the learned sampling patterns for $R=5$ (a) and $R=10$ (b): Top row shows example slices from the knee and brain datasets, middle row shows the learned sampling patterns, and the bottom row shows the corresponding PSFs in logarithmic scale. Patterns optimized for the knee and brain datasets have different characteristics, affecting the overall sample distribution and PSFs. Pattern optimization produces diffused sidelobes depending on the anatomy. The patterns differ due to different energy spectrums of the images in the two datasets. Note that only the ground truth knee data was acquired in corner-cut fashion, which is also reflected in the optimized patterns.}
\label{fig:anatomy}
\end{figure}

We test the dependence on the training sets by comparing the learned patterns from the brain and knee datasets. Fig. \ref{fig:anatomy} shows the sampling patterns along with the PSFs for $R=5$ and $R=10$ with the noise variance $\sigma=0.0001$. The sampling patterns show different characteristics due to the differences in image energy spectra, as well as image and k-space supports of these two datasets. 
Appendix~\ref{appendix:cross_anatomy} compares reconstruction quality of the brain-optimized pattern on the knee test set and demonstrates the anatomy-specific nature of optimized sampling patterns from an image quality perspective.

\subsection{Impact of Noise} \label{sec:impact_of_noise}
\begin{figure*}[h!]
\centerline{\includegraphics[width=0.95\linewidth]{Figures/alkan7.png}}
\caption{Effect of measurement noise on sampling density and k-space extent for an acceleration factor $R=10$ on the knee dataset. The first five columns show the learned sampling patterns for the noise levels $\sigma=\{1e-4, 1e-3, 1e-2, 3e-2, 1e-1\}$, and the final column is the Voronoi cell area distribution. As the measurement noise increases, the high-frequency samples become mostly noise-corrupted. The noise amount directly affects the k-space extent of the optimized samples. The Voronoi cell area distribution also indicates that as the noise amount increases, the learned patterns have more low-frequency samples. The result for $\sigma=1e-4$ is obtained with uniform initialization and all of the other results are obtained with the uniform small disc initialization.}
\label{fig:noise}
\end{figure*}
We study the effect of measurement noise on the learned sampling patterns by changing the standard deviation $\sigma$ of the complex Gaussian noise $\epsilon$ in Equation \ref{eq:1}. We consider five different noise levels: $\sigma=\{1e-4, 1e-3, 1e-2, 3e-2, 1e-1\}$ with an acceleration factor $R=10$ on the knee dataset and compare the Voronoi cell area distributions in Fig. \ref{fig:noise}. As the signal-to-noise ratio (SNR) decreases, the optimal sampling patterns concentrate around the center since collecting highly noise-corrupted outer k-space would degrade the reconstruction quality. This phenomenon was also reported in \cite{aggarwal2020j}.

\subsection{Effect of Coil Sensitivities}
\begin{figure}[h!]
\centerline{\includegraphics[width=1.07\columnwidth]{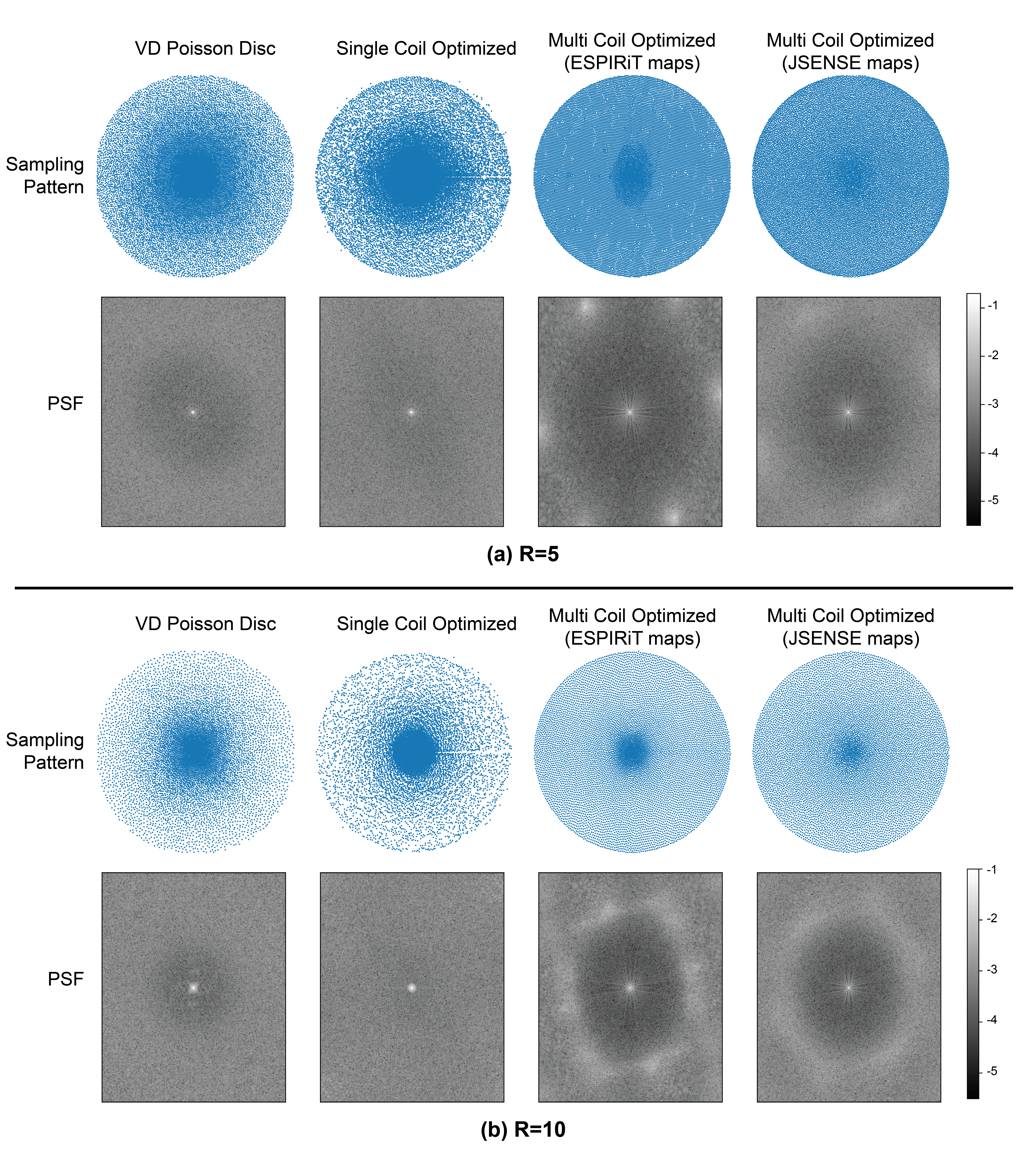}}
\caption{Effect of coil sensitivities on the learned sampling patterns for $R=5$ (a) and $R=10$ (b). Top rows show the patterns and the bottom rows show the PSFs in logarithmic scale. The single-coil and multi-coil optimized patterns are remarkably different. Single-coil patterns resemble the random Variable Density pattern. Multi-coil patterns are more uniformly distributed and their PSF sidelobes are suppressed at the center. Overall, optimized samples adapt to coil sensitivity maps during training.}
\label{fig:coilsens}
\end{figure}

In order to assess the influence of coil sensitivities on the learned sampling patterns, we run experiments that compare the patterns from multi-coil and emulated single-coil datasets. For single coil experiments, we treat the SENSE combined images as ground truth as in multi-coil experiments but remove the coil sensitivity dependence in the forward signal model in Eq. \ref{eq:1} and Eq. \ref{eq:6}. For the multi-coil scenario, we compare the resulting patterns trained with ESPIRiT and JSENSE estimated maps. Fig. \ref{fig:coilsens} demonstrates the single-coil and multi-coil optimized patterns along with their PSFs for $R=5$ and $R=10$. Single-coil optimized patterns conform more to a random variable density distribution, whereas the samples in multi-coil patterns are more structured and they are distributed more uniformly. Moreover, the PSFs of the multi-coil patterns have much smaller sidelobe amplitudes at the center, and larger sidelobes towards the outside of the FOV. Additionally, patterns learned with JSENSE maps are slightly different than the ones learned with ESPIRiT maps, signifying the dependence on the sensitivity maps in the training dataset. Appendix~\ref{appendix:cross_sensitivity} further explores this comparison from an image quality perspective.

\begin{figure}[h!]
\centerline{\includegraphics[width=0.8\columnwidth]{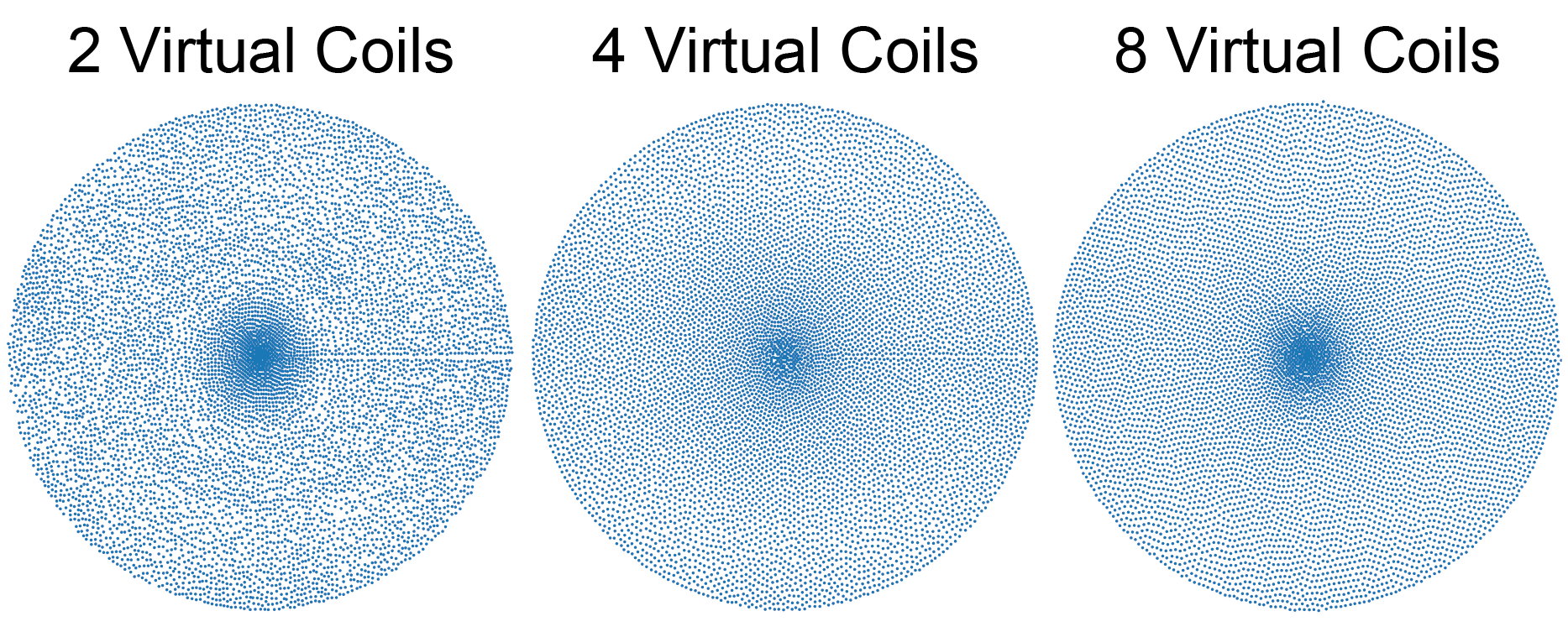}}
\caption{Optimized sampling patterns on the knee dataset using different numbers of virtual coils for $R=10$. The coils were compressed by SVD coil compression. As the number of compressed coils decreases, the learned patterns get denser at the center and start having more randomness in high frequencies.}
\label{fig:coil_compress}
\end{figure}

Fig. \ref{fig:coil_compress} illustrates the changes in sampling patterns when different numbers of virtual coils are used after coil compression. As the number of virtual coils decreases, the sampling density in the low frequencies increases and sampling in the high frequencies becomes more random. The patterns start resembling the variable density random patterns as in the single-coil scenario in Fig. \ref{fig:coilsens} since the reconstruction problem becomes less dependent on the coil information.

\subsection{Prospective Acquisitions}

Since we use a continuous representation for the k-space coordinates, the phase encodes we obtain via AutoSamp do not necessarily fall into Cartesian grid locations. These patterns can be generated in the MRI system for the 3D imaging scenario by scaling gradient waveforms appropriately. However, it is also possible to get a fully-Cartesian sampling pattern by rounding off each sample location to the nearest Cartesian grid point. This discretization approach makes the pulse sequence implementation easier. In Table \ref{table:metrics_discretized}, we show that this is a viable strategy since the reconstruction network trained for the non-discretized sampling pattern performs similarly on the data obtained with the discretized sampling patterns, especially for high acceleration factors. For our prospective acquisitions, we implement the discretized version of the AutoSamp optimized patterns.

\begin{table}[h!]
\caption{Effect of rounding off samples to nearest Cartesian grid point}
\label{table:metrics_discretized}
\centering
\begin{tabular}{cccc}
    \toprule
    Acceleration ($R$) & Sampling Pattern & PSNR & SSIM \\
    \midrule
    \multirow{2}{*}{5x} & Non-Discretized & 44.24 & 0.988 \\
     & Discretized & {43.61} & {0.987} \\\midrule
    \multirow{2}{*}{10x} & Non-Discretized & 35.95 & 0.909 \\
     & Discretized & {35.13} & {0.901} \\\midrule
    \multirow{2}{*}{15x} & Non-Discretized & 33.87 & 0.859 \\
     & Discretized & {33.93} & 0.859 \\\midrule
    \multirow{2}{*}{20x} & Non-Discretized & 33.03 & {0.817} \\
     & Discretized & {33.00} & {0.816} \\\midrule
    \end{tabular}
\end{table}
\begin{figure*}[h!]
\centerline{\includegraphics[width=0.85\linewidth]{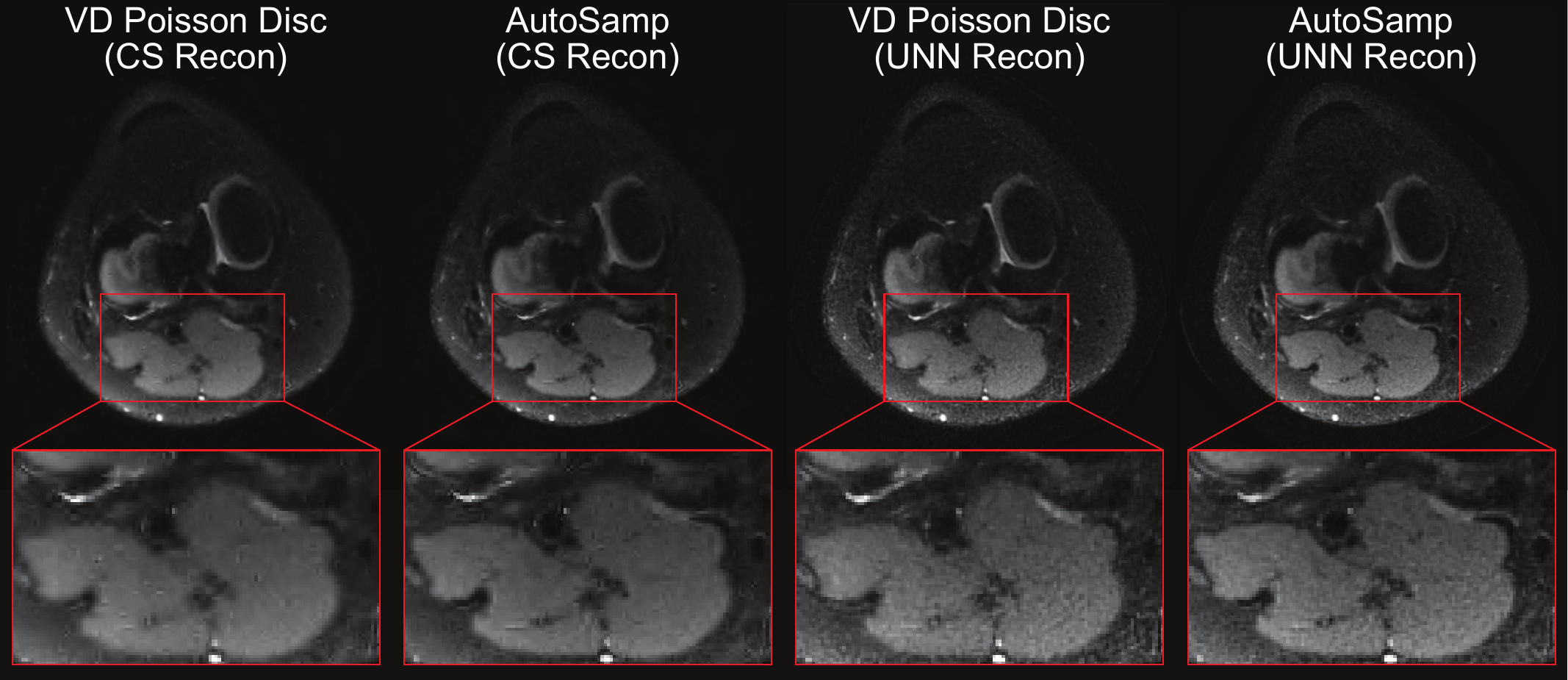}}
\caption{Representative results for the prospective acquisitions for $R=20\times$ acceleration with the VD Poisson Disc and AutoSamp-optimized patterns. CS and UNN reconstructions for each pattern are shown. Optimized pattern improves the reconstruction quality for both CS and UNN reconstructions by improving sharpness and reducing noise level. Zoomed in regions illustrate that our method preserves structural details better.}
\label{fig:prospective_20x}
\end{figure*}
\begin{figure*}[h!]
\centerline{\includegraphics[width=\linewidth]{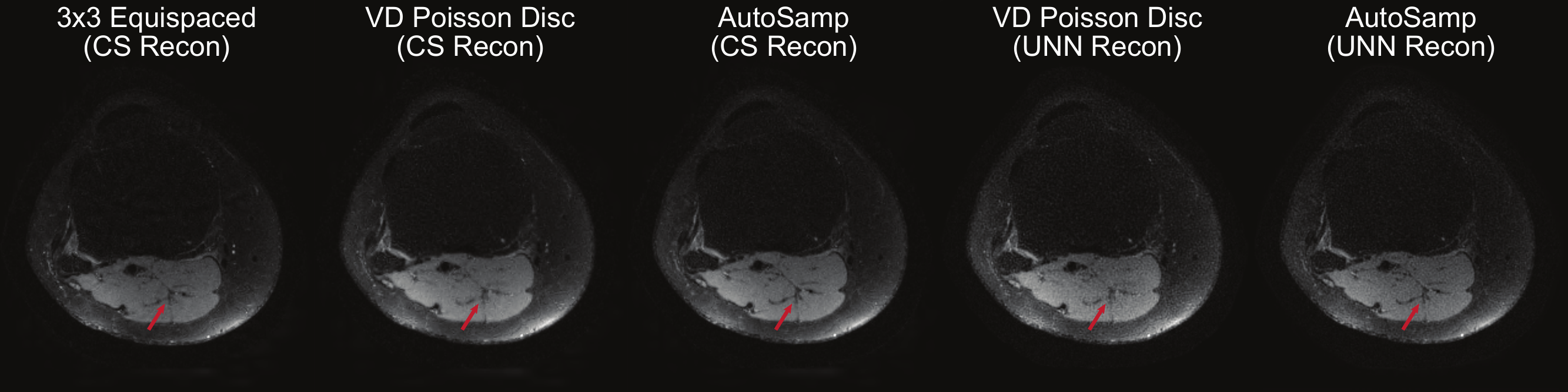}}
\caption{Representative results for the prospective acquisitions for $R=10\times$ acceleration. CS and UNN reconstructions are shown for AutoSamp-optimized and VD Poisson Disc patterns. Vendor's maximum allowed default undersampling scheme, 3x3 equispaced, is also shown with CS reconstruction. Optimized pattern improves the sharpness of reconstructed images.}
\label{fig:prospective_10x}
\end{figure*}
Fig. \ref{fig:prospective_20x} and Fig. \ref{fig:prospective_10x} show representative CS and UNN reconstruction results for the prospective acquisitions with AutoSamp-optimized and VD Poisson Disc patterns. We observe that the optimized pattern improves the reconstruction quality for both CS and UNN reconstructions. In particular, UNN reconstruction in Fig. \ref{fig:prospective_20x} using the optimized pattern for $R=20$ produces much sharper images and preserves the structural details better than the VD Poisson Disc pattern. Furthermore, the noise level is smaller on the reconstructions with the optimized pattern.
\section{Discussion}
We introduced a variational information maximization framework that enables joint optimization of MRI data sampling and reconstruction. We obtained a loss function that depends on both the sample locations and the reconstruction function parameters by constructing a lower bound on the mutual information between the image domain and noisy k-space representations. Representing the phase encoding locations as continuous variables and utilizing NUFFT allowed us to use gradients and backpropagation for joint optimization.

Our experiments showed the benefit of jointly learning the sampling pattern and reconstruction network. Across a wide range of acceleration factors, the patterns optimized with our proposed AutoSamp algorithm provided improved image quality metrics compared to the traditionally used and other DL-based sample optimization methods. We observed that the improvements are higher for lower acceleration factors which have more degrees of freedom for pattern design.

We implemented AutoSamp-optimized sampling patterns prospectively with 3D FSE CUBE sequences for proton density weighted knee acquisitions. Our prospective evaluations demonstrated the improved image quality and sharpness obtained with our optimized patterns, especially for UNN reconstruction. Considering the wide clinical use of 3D FSE acquisitions, our sampling optimization framework bears the potential to accelerate the 3D MRI protocols for other body regions and contrasts.

We analyzed the sampling patterns with respect to changes in acceleration factor, measurement noise, anatomy and coil sensitivities. Even though the learned sampling patterns adjust the k-space density according to the acceleration factor, all of the learned patterns have common features such as non-clustered samples and more uniform density across k-space. Similarly, the measurement noise affects the sampling density and k-space extent. Moreover, the patterns learned from different anatomies are different due to different energy spectra. Finally, we showed that the learned sampling patterns inherently depend on the coil sensitivities, requiring completely different patterns for single-coil and multi-coil scenarios. Overall, our results indicate that the learned sampling patterns adapt to each factor we investigated by adjusting the sampling density, k-space coverage and point spread functions.

Our proposed method was designed to optimize patterns for unrolled neural network reconstruction, however we found that the learned patterns also improve the reconstruction quality for compressed sensing reconstruction. This can be attributed to the fact that the PGD based UNN deployed in this work mimics the optimization process for CS solutions. Future work could focus on making theoretical or empirical connections between Wavelet domain sparsity offered by CS algorithms and the implicit sparsity enforced by the UNNs. Such connections could provide more insights and enhance the interpretability of learned sampling patterns and UNN reconstructions.

A natural extension could be using CS reconstruction as the decoder. In our early experiments, we found that using CS as a decoder resulted in instabilities or vanishing gradient problems during optimization, causing the patterns to get stuck at a local minima. This observation aligns with the findings outlined in \cite{gossard2022spurious}, which offers a mathematical perspective on the challenges encountered when optimizing sampling patterns with linear non-parameterized reconstruction methods. Conversely, using a fully-parameterized UNN as the decoder aligns better with the variational information maximization framework and it resulted in smoother optimization with improved results.

Quantitative results in Table \ref{table:metrics} indicate that CS reconstruction performance is higher than of UNN for $R=5$ and $R=10$. Increasing the number of unrolled iterations or using deeper proximal blocks can further improve the result of UNN reconstruction at the cost of a higher GPU memory budget. However, our main focus in this work is the improvement obtained via optimized sampling patterns for the same type of reconstruction.

Certain design choices still need to be made for efficient optimization of sampling patterns. Our hyperparameter search revealed that keeping a lower learning rate ($0.001$) for neural network optimization and using a higher learning rate ($0.01$) for the sample pattern optimization yields faster convergence and prevents instabilities during training. In addition, the initial sampling pattern choice affects the optimization progress. In our experiments, we found that the uniform or Poisson Disc initialization usually provides faster convergence and better final reconstruction quality. When the SNR is low, starting from a uniform small disc initialization produces better results since the samples are more concentrated at low spatial frequencies as described in Sec. \ref{sec:impact_of_noise}. Note that due to the highly non-convex nature of the joint optimization problem, the final patterns obtained with different initializations do not necessarily coincide. Therefore, sampling pattern initialization should be treated as a hyperparameter and the best initialization should be chosen based on the validation set results.

In our study, we specifically focused on the 3D acquisition setting with Cartesian readouts. AutoSamp can also be used to design parameterized 2D and 3D trajectories by optimizing spiral, cone interleaf and radial spoke angles. Additionally, our method can be extended for the design of continuous trajectories by incorporating MRI system constraints during optimization. Such approaches would additionally require resolving trajectory errors due to eddy currents, which can be addressed by employing trajectory measurement methods \cite{wang2022b}.

Furthermore, joint optimization via AutoSamp can also be applied to other imaging scenarios such as dynamic imaging. Patterns that explore both the temporal and spatial redundancies effectively in k-t space can further improve the reconstruction quality and provide more insights. Additionally, multi-contrast reconstruction problem which jointly reconstructs multiple clinical contrasts from accelerated MRI acquisitions can benefit from the careful design of sampling patterns. The acquisitions share anatomical information across the imaging contrasts and sampling patterns can exploit spatial redundancies by considering information from the k-space of other contrasts. Future work will consider k-t and multicontrast pattern optimization scenarios.

We observed that the optimized patterns resemble hexagonal packing, especially for $R=5$ and $R=10$. Prior works have explored hexagonal sampling patterns with uniform sampling densities across k-space \cite{engel2021t,saranathan2007anthem,breuer2006controlled}. An interesting future research direction is to leverage the sampling statistics and design insights from AutoSamp to generate hexagonal patterns parameterized according to specified k-space densities. While this approach simplifies pattern design, it might neglect potential benefits gained from incorporating anatomy-specific features, coil sensitivity information, and noise levels into the sampling patterns.

On the opposite end of the spectrum, factors influencing the sampling patterns can be integrated directly into the training process to facilitate the learning of adaptive, subject-specific sampling patterns. Coil sensitivities, noise levels, and anatomical boundaries can be estimated from fast calibration scans. The sampling patterns can be adapted at the test time based on the information gathered from the calibration scans. Future work could explore utilizing the insights derived from AutoSamp to learn adaptive sampling patterns.

Our framework additionally allows incorporating other MR physics related artifacts beyond the Gaussian measurement noise such as motion, off-resonance and relaxation effects. These artifacts can be simulated as a part of the forward model during the training phase to find the optimal sampling locations in the presence of such effects. Potential future work could focus on generalizing the sampling pattern optimization to these scenarios.
\section{Conclusion}
In this work, we presented AutoSamp, a data-driven framework based on variational information maximization that enables joint optimization of MRI data sampling and reconstruction. Experiments on public datasets and prospective acquisitions demonstrate the effectiveness of our method across a wide range of acceleration factors. Empirical analysis reveal the dependence of the learned sampling patterns on acceleration factor, measurement noise, underlying anatomy, and coil sensitivities.
\appendices
\section*{Appendix}
\renewcommand{\thesubsection}{\Alph{subsection}}
\subsection{Zoomed-in Views of Sampling Patterns}
\begin{figure}[h!]
\centerline{\includegraphics[width=\columnwidth]{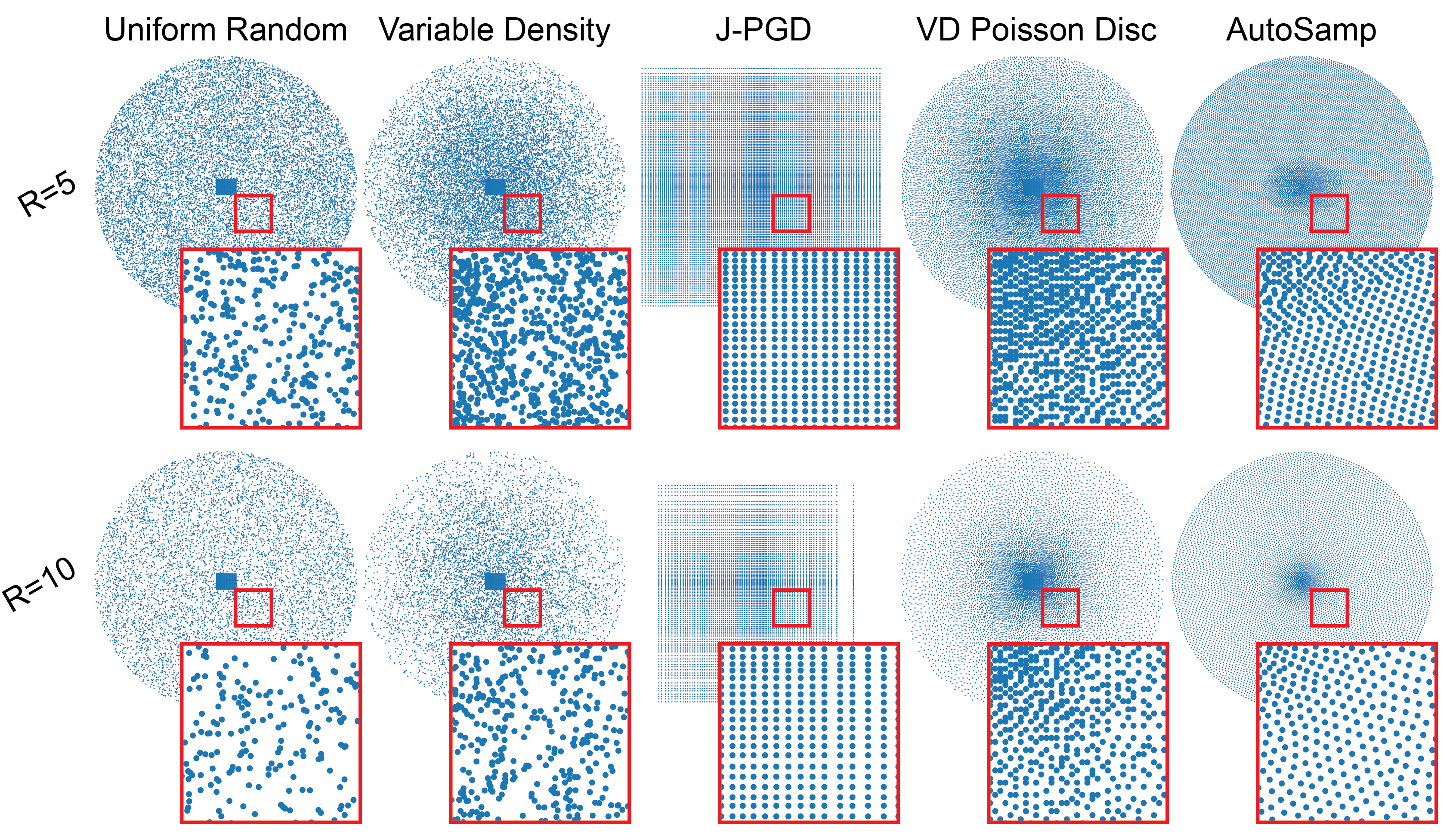}}
\caption{Zoom-in views of the sampling patterns for $R=\{5,10\}$.}
\label{fig:zoomin}
\end{figure}

Fig. \ref{fig:zoomin} displays the zoomed-in views of the sampling patterns used in our knee dataset experiments for $R=5$ and $R=10$.

\subsection{Impact of Joint Optimization Strategy}
\label{appendix:opt_strategy}
We compared joint optimization performance with optimizing only the reconstruction network ($\theta$-only) and sequentially optimizing sampling and reconstruction ($\theta,\phi$-sequential). For the sequential optimization, we kept the reconstruction network weights ($\theta$) the same as the $\theta$-only strategy, and optimized the sampling pattern for 100 more epochs. While joint optimization demonstrate the combined contribution of sampling and reconstruction, $\theta$-only and $\theta,\phi$-sequential optimization strategies show the individual contributions.

We present the impact of optimization strategies in Table \ref{table:optimization_strategy}. Our results indicate the the joint optimization strategy we use in AutoSamp provides the highest image quality metrics, highlighting the benefits of simultaneously optimizing sampling and reconstruction for UNN reconstructions.

\begin{table}[h!]
\caption{Impact of Optimization Strategies on Knee Test Set}
\label{table:optimization_strategy}
\centering
\resizebox{\columnwidth}{!}{
\begin{tabular}{cccc|cc}
    \toprule
    \multirow{2}{*}{Strategy} & \multirow{2}{*}{Sampling Pattern} & \multicolumn{2}{c}{$R=10$} & \multicolumn{2}{c}{$R=20$} \\
    & & PSNR & SSIM & PSNR & SSIM \\
    \midrule
    $\theta$ only & Variable Density & 33.82 & 0.862 & 31.97 & 0.781 \\
    $\theta$ only & VD Poisson Disc & 33.94 & 0.872 & 32.33 & 0.790 \\
    $\theta, \phi$ sequential & Variable Density & 34.88 & 0.885 & 32.45 & 0.791 \\
    $\theta, \phi$ sequential & VD Poisson Disc & 33.98 & 0.872 & 32.51 & 0.799 \\
    $\theta, \phi$ joint & AutoSamp & \bfseries35.95 & \bfseries0.909 & \bfseries33.03 & \bfseries0.817 \\\midrule
    \end{tabular}
}
\end{table}

\subsection{Cross Anatomy Evaluation}
\label{appendix:cross_anatomy}
We test the reconstruction performance of the brain-optimized pattern on the knee test set in Table \ref{table:metrics_braintoknee}. We compared the patterns with the same number of samples and used AutoSamp's UNN trained on the knee dataset for reconstruction. Additionally, we fine-tuned the brain dataset's UNN on the knee dataset (while keeping the brain pattern fixed) and reported its performance. We observe that reconstructions with the brain-optimized patterns have much lower image quality on knees for all acceleration factors, supporting the observation that the learned sampling patterns are anatomy-specific.

\begin{table}[h!]
\caption{Performance of the brain-optimized pattern on the knee test set}
\label{table:metrics_braintoknee}
\centering
\resizebox{\columnwidth}{!}{
\begin{tabular}{ccccc}
    \toprule
    \multirow{2}{*}{Acceleration} & \multirow{2}{*}{Sampling Pattern} & \multirow{2}{*}{Reconstruction} & \multicolumn{2}{c}{Metrics (on knees)} \\
    \cmidrule{4-5}
    & & & PSNR & SSIM \\
    \midrule
    \multirow{4}{*}{5x} & AutoSamp Knee & AutoSamp Knee & \bfseries44.24 & \bfseries0.988 \\
     & AutoSamp Brain & AutoSamp Knee & 25.67 & 0.666 \\
     & \multirow{2}{*}{AutoSamp Brain} & AutoSamp Brain & \multirow{2}{*}{39.00} & \multirow{2}{*}{0.955} \\
     & & (fine-tuned on knee) & & \\\midrule
    \multirow{4}{*}{10x} & AutoSamp Knee & AutoSamp Knee & \bfseries35.95 & \bfseries0.909 \\
     & AutoSamp Brain & AutoSamp Knee & 27.78 & 0.790 \\
     & \multirow{2}{*}{AutoSamp Brain} & AutoSamp Brain & \multirow{2}{*}{32.05} & \multirow{2}{*}{0.822} \\
     & & (fine-tuned on knee) & & \\\midrule
    \multirow{4}{*}{15x} & AutoSamp Knee & AutoSamp Knee & \bfseries33.87 & \bfseries0.859 \\
     & AutoSamp Brain & AutoSamp Knee & 21.11 & 0.653 \\
     & \multirow{2}{*}{AutoSamp Brain} & AutoSamp Brain & \multirow{2}{*}{31.38} & \multirow{2}{*}{0.793} \\
     & & (fine-tuned on knee) & & \\\midrule
    \multirow{4}{*}{20x} & AutoSamp Knee & AutoSamp Knee & \bfseries33.03 & \bfseries0.817 \\
     & AutoSamp Brain & AutoSamp Knee & 29.31 & 0.777 \\
     & \multirow{2}{*}{AutoSamp Brain} & AutoSamp Brain & \multirow{2}{*}{30.25} & \multirow{2}{*}{0.777} \\
     & & (fine-tuned on knee) & & \\\midrule
    \end{tabular}
}
\end{table}

\subsection{Reconstruction Quality Assessment of JSENSE and ESPIRiT Optimized Patterns}
\label{appendix:cross_sensitivity}

We evaluate the reconstruction performance of the JSENSE optimized sampling patterns for reconstructions using ESPIRiT sensitivity maps in Table \ref{table:metrics_jsensetoespirit}. Our results indicate that the image quality metrics decrease slightly when JSENSE optimized pattern is used for UNN reconstruction. The reconstruction quality between the patterns is comparable when using CS reconstruction. The performance of deep learning based reconstruction models depends on the training set, including the coil sensitivities. Consequently, we observe a slight decrease in UNN reconstruction quality in this cross-sensitivity map scenario.

\begin{table}[h!]
\caption{Performance of jsense-optimized patterns on the knee test set with espirit maps}
\label{table:metrics_jsensetoespirit}
\centering
\resizebox{\columnwidth}{!}{
\begin{tabular}{cccccc}
    \toprule
    \multirow{2}{*}{Reconstruction} & \multirow{2}{*}{Sampling Pattern} & \multicolumn{4}{c}{SSIM} \\
    & & $R=5$ & $R=10$ & $R=15$ & $R=20$ \\
    \midrule
    \multirow{2}{*}{UNN} & ESPIRiT optimized & 0.988 & 0.909 & 0.859 & 0.817 \\
                         & JSENSE optimized & 0.975 & 0.894 & 0.861 & 0.812 \\\midrule
    \multirow{2}{*}{CS}  & ESPIRiT optimized & 0.993 & 0.927 & 0.868 & 0.831 \\
                         & JSENSE optimized & 0.988 & 0.919 & 0.861 & 0.827 \\\midrule
    \end{tabular}
}
\end{table}

\bibliographystyle{ieeetr}
\bibliography{references}

\begin{thebibliography}{10}

\bibitem{lustig2007sparse}
M.~Lustig, D.~Donoho, and J.~M. Pauly, ``Sparse {MRI}: The application of compressed sensing for rapid {MR} imaging,'' {\em Magnetic Resonance in Medicine: An Official Journal of the International Society for Magnetic Resonance in Medicine}, vol.~58, no.~6, pp.~1182--1195, 2007.

\bibitem{lustig2010spirit}
M.~Lustig and J.~M. Pauly, ``{SPIRiT}: iterative self-consistent parallel imaging reconstruction from arbitrary k-space,'' {\em Magnetic resonance in medicine}, vol.~64, no.~2, pp.~457--471, 2010.

\bibitem{candes2006stable}
E.~J. Candes, J.~K. Romberg, and T.~Tao, ``Stable signal recovery from incomplete and inaccurate measurements,'' {\em Communications on Pure and Applied Mathematics: A Journal Issued by the Courant Institute of Mathematical Sciences}, vol.~59, no.~8, pp.~1207--1223, 2006.

\bibitem{donoho2006compressed}
D.~L. Donoho, ``Compressed sensing,'' {\em IEEE Transactions on information theory}, vol.~52, no.~4, pp.~1289--1306, 2006.

\bibitem{bridson2007fast}
R.~Bridson, ``Fast {Poisson} disk sampling in arbitrary dimensions.,'' {\em SIGGRAPH sketches}, vol.~10, no.~1, p.~1, 2007.

\bibitem{vasanawala2011practical}
S.~Vasanawala, M.~Murphy, M.~T. Alley, P.~Lai, K.~Keutzer, J.~M. Pauly, and M.~Lustig, ``Practical parallel imaging compressed sensing {MRI}: Summary of two years of experience in accelerating body {MRI} of pediatric patients,'' in {\em 2011 ieee international symposium on biomedical imaging: From nano to macro}, pp.~1039--1043, IEEE, 2011.

\bibitem{knoll2011adapted}
F.~Knoll, C.~Clason, C.~Diwoky, and R.~Stollberger, ``Adapted random sampling patterns for accelerated {MRI},'' {\em Magnetic resonance materials in physics, biology and medicine}, vol.~24, pp.~43--50, 2011.

\bibitem{seeger2010optimization}
M.~Seeger, H.~Nickisch, R.~Pohmann, and B.~Sch{\"o}lkopf, ``Optimization of k-space trajectories for compressed sensing by {Bayesian} experimental design,'' {\em Magnetic Resonance in Medicine: An Official Journal of the International Society for Magnetic Resonance in Medicine}, vol.~63, no.~1, pp.~116--126, 2010.

\bibitem{gozcu2018learning}
B.~G{\"o}zc{\"u}, R.~K. Mahabadi, Y.-H. Li, E.~Il{\i}cak, T.~Cukur, J.~Scarlett, and V.~Cevher, ``Learning-based compressive {MRI},'' {\em IEEE transactions on medical imaging}, vol.~37, no.~6, pp.~1394--1406, 2018.

\bibitem{sanchez2020scalable}
T.~Sanchez, B.~G{\"o}zc{\"u}, R.~B. van Heeswijk, A.~Eftekhari, E.~Il{\i}cak, T.~{\c{C}}ukur, and V.~Cevher, ``Scalable learning-based sampling optimization for compressive dynamic {MRI},'' in {\em ICASSP 2020-2020 IEEE International Conference on Acoustics, Speech and Signal Processing (ICASSP)}, pp.~8584--8588, IEEE, 2020.

\bibitem{zibetti2021fast}
M.~V. Zibetti, G.~T. Herman, and R.~R. Regatte, ``Fast data-driven learning of parallel {MRI} sampling patterns for large scale problems,'' {\em Scientific Reports}, vol.~11, no.~1, p.~19312, 2021.

\bibitem{haldar2019oedipus}
J.~P. Haldar and D.~Kim, ``{OEDIPUS}: An experiment design framework for sparsity-constrained {MRI},'' {\em IEEE transactions on medical imaging}, vol.~38, no.~7, pp.~1545--1558, 2019.

\bibitem{lazarus2019sparkling}
C.~Lazarus, P.~Weiss, N.~Chauffert, F.~Mauconduit, L.~El~Gueddari, C.~Destrieux, I.~Zemmoura, A.~Vignaud, and P.~Ciuciu, ``{SPARKLING}: variable-density k-space filling curves for accelerated {T2*-weighted} {MRI},'' {\em Magnetic resonance in medicine}, vol.~81, no.~6, pp.~3643--3661, 2019.

\bibitem{liang2020deep}
D.~Liang, J.~Cheng, Z.~Ke, and L.~Ying, ``Deep magnetic resonance image reconstruction: Inverse problems meet neural networks,'' {\em IEEE Signal Processing Magazine}, vol.~37, no.~1, pp.~141--151, 2020.

\bibitem{hammernik2023physics}
K.~Hammernik, T.~K{\"u}stner, B.~Yaman, Z.~Huang, D.~Rueckert, F.~Knoll, and M.~Ak{\c{c}}akaya, ``Physics-driven deep learning for computational magnetic resonance imaging: Combining physics and machine learning for improved medical imaging,'' {\em IEEE signal processing magazine}, vol.~40, no.~1, pp.~98--114, 2023.

\bibitem{knoll2020deep}
F.~Knoll, K.~Hammernik, C.~Zhang, S.~Moeller, T.~Pock, D.~K. Sodickson, and M.~Akcakaya, ``Deep-learning methods for parallel magnetic resonance imaging reconstruction: A survey of the current approaches, trends, and issues,'' {\em IEEE signal processing magazine}, vol.~37, no.~1, pp.~128--140, 2020.

\bibitem{wang2016accelerating}
S.~Wang, Z.~Su, L.~Ying, X.~Peng, S.~Zhu, F.~Liang, D.~Feng, and D.~Liang, ``Accelerating magnetic resonance imaging via deep learning,'' in {\em 2016 IEEE 13th international symposium on biomedical imaging (ISBI)}, pp.~514--517, IEEE, 2016.

\bibitem{hyun2018deep}
C.~M. Hyun, H.~P. Kim, S.~M. Lee, S.~Lee, and J.~K. Seo, ``Deep learning for undersampled {MRI} reconstruction,'' {\em Physics in Medicine \& Biology}, vol.~63, no.~13, p.~135007, 2018.

\bibitem{lee2018deep}
D.~Lee, J.~Yoo, S.~Tak, and J.~C. Ye, ``Deep residual learning for accelerated {MRI} using magnitude and phase networks,'' {\em IEEE Transactions on Biomedical Engineering}, vol.~65, no.~9, pp.~1985--1995, 2018.

\bibitem{han2019k}
Y.~Han, L.~Sunwoo, and J.~C. Ye, ``$k$-space deep learning for accelerated {MRI},'' {\em IEEE transactions on medical imaging}, vol.~39, no.~2, pp.~377--386, 2019.

\bibitem{eo2018kiki}
T.~Eo, Y.~Jun, T.~Kim, J.~Jang, H.-J. Lee, and D.~Hwang, ``{KIKI-net}: cross-domain convolutional neural networks for reconstructing undersampled magnetic resonance images,'' {\em Magnetic resonance in medicine}, vol.~80, no.~5, pp.~2188--2201, 2018.

\bibitem{akccakaya2019scan}
M.~Ak{\c{c}}akaya, S.~Moeller, S.~Weing{\"a}rtner, and K.~U{\u{g}}urbil, ``Scan-specific robust artificial-neural-networks for k-space interpolation ({RAKI}) reconstruction: Database-free deep learning for fast imaging,'' {\em Magnetic resonance in medicine}, vol.~81, no.~1, pp.~439--453, 2019.

\bibitem{zhu2018image}
B.~Zhu, J.~Z. Liu, S.~F. Cauley, B.~R. Rosen, and M.~S. Rosen, ``Image reconstruction by domain-transform manifold learning,'' {\em Nature}, vol.~555, no.~7697, pp.~487--492, 2018.

\bibitem{hammernik2018learning}
K.~Hammernik, T.~Klatzer, E.~Kobler, M.~P. Recht, D.~K. Sodickson, T.~Pock, and F.~Knoll, ``Learning a variational network for reconstruction of accelerated {MRI} data,'' {\em Magnetic resonance in medicine}, vol.~79, no.~6, pp.~3055--3071, 2018.

\bibitem{schlemper2017deep}
J.~Schlemper, J.~Caballero, J.~V. Hajnal, A.~Price, and D.~Rueckert, ``A deep cascade of convolutional neural networks for {MR} image reconstruction,'' in {\em Information Processing in Medical Imaging: 25th International Conference, IPMI 2017, Boone, NC, USA, June 25-30, 2017, Proceedings 25}, pp.~647--658, Springer, 2017.

\bibitem{schlemper2018deep}
J.~Schlemper, J.~Caballero, J.~V. Hajnal, A.~N. Price, and D.~Rueckert, ``A deep cascade of convolutional neural networks for dynamic {MR} image reconstruction,'' {\em IEEE Transactions on Medical Imaging}, vol.~37, no.~2, pp.~491--503, 2018.

\bibitem{yang2018admm}
Y.~Yang, J.~Sun, H.~Li, and Z.~Xu, ``{ADMM-CSNet}: A deep learning approach for image compressive sensing,'' {\em IEEE transactions on pattern analysis and machine intelligence}, vol.~42, no.~3, pp.~521--538, 2018.

\bibitem{zhang2018ista}
J.~Zhang and B.~Ghanem, ``{ISTA-Net}: Interpretable optimization-inspired deep network for image compressive sensing,'' in {\em Proceedings of the IEEE conference on computer vision and pattern recognition}, pp.~1828--1837, 2018.

\bibitem{cheng2019model}
J.~Cheng, H.~Wang, L.~Ying, and D.~Liang, ``Model learning: Primal dual networks for fast {MR} imaging,'' in {\em Medical Image Computing and Computer Assisted Intervention--MICCAI 2019: 22nd International Conference, Shenzhen, China, October 13--17, 2019, Proceedings, Part III 22}, pp.~21--29, Springer, 2019.

\bibitem{aggarwal2018modl}
H.~K. Aggarwal, M.~P. Mani, and M.~Jacob, ``{MoDL}: Model-based deep learning architecture for inverse problems,'' {\em IEEE transactions on medical imaging}, vol.~38, no.~2, pp.~394--405, 2018.

\bibitem{cheng2018highly}
J.~Y. Cheng, F.~Chen, M.~T. Alley, J.~M. Pauly, and S.~S. Vasanawala, ``Highly scalable image reconstruction using deep neural networks with bandpass filtering,'' {\em arXiv preprint arXiv:1805.03300}, 2018.

\bibitem{sandino2020compressed}
C.~M. Sandino, J.~Y. Cheng, F.~Chen, M.~Mardani, J.~M. Pauly, and S.~S. Vasanawala, ``Compressed sensing: From research to clinical practice with deep neural networks: Shortening scan times for magnetic resonance imaging,'' {\em IEEE signal processing magazine}, vol.~37, no.~1, pp.~117--127, 2020.

\bibitem{yang2017dagan}
G.~Yang, S.~Yu, H.~Dong, G.~Slabaugh, P.~L. Dragotti, X.~Ye, F.~Liu, S.~Arridge, J.~Keegan, Y.~Guo, {\em et~al.}, ``{DAGAN}: deep de-aliasing generative adversarial networks for fast compressed sensing {MRI} reconstruction,'' {\em IEEE transactions on medical imaging}, vol.~37, no.~6, pp.~1310--1321, 2017.

\bibitem{quan2018compressed}
T.~M. Quan, T.~Nguyen-Duc, and W.-K. Jeong, ``Compressed sensing {MRI} reconstruction using a generative adversarial network with a cyclic loss,'' {\em IEEE transactions on medical imaging}, vol.~37, no.~6, pp.~1488--1497, 2018.

\bibitem{mardani2018deep}
M.~Mardani, E.~Gong, J.~Y. Cheng, S.~S. Vasanawala, G.~Zaharchuk, L.~Xing, and J.~M. Pauly, ``Deep generative adversarial neural networks for compressive sensing {MRI},'' {\em IEEE transactions on medical imaging}, vol.~38, no.~1, pp.~167--179, 2018.

\bibitem{darestani2021accelerated}
M.~Z. Darestani and R.~Heckel, ``Accelerated {MRI} with un-trained neural networks,'' {\em IEEE Transactions on Computational Imaging}, vol.~7, pp.~724--733, 2021.

\bibitem{yoo2021time}
J.~Yoo, K.~H. Jin, H.~Gupta, J.~Yerly, M.~Stuber, and M.~Unser, ``Time-dependent deep image prior for dynamic {MRI},'' {\em IEEE Transactions on Medical Imaging}, vol.~40, no.~12, pp.~3337--3348, 2021.

\bibitem{jin2019self}
K.~H. Jin, M.~Unser, and K.~M. Yi, ``Self-supervised deep active accelerated {MRI},'' {\em arXiv preprint arXiv:1901.04547}, 2019.

\bibitem{zhang2019reducing}
Z.~Zhang, A.~Romero, M.~J. Muckley, P.~Vincent, L.~Yang, and M.~Drozdzal, ``Reducing uncertainty in undersampled {MRI} reconstruction with active acquisition,'' in {\em Proceedings of the IEEE/CVF Conference on Computer Vision and Pattern Recognition}, pp.~2049--2058, 2019.

\bibitem{pineda2020active}
L.~Pineda, S.~Basu, A.~Romero, R.~Calandra, and M.~Drozdzal, ``Active {MR} k-space sampling with reinforcement learning,'' in {\em Medical Image Computing and Computer Assisted Intervention--MICCAI 2020: 23rd International Conference, Lima, Peru, October 4--8, 2020, Proceedings, Part II 23}, pp.~23--33, Springer, 2020.

\bibitem{bakker2020experimental}
T.~Bakker, H.~van Hoof, and M.~Welling, ``Experimental design for {MRI} by greedy policy search,'' {\em Advances in Neural Information Processing Systems}, vol.~33, pp.~18954--18966, 2020.

\bibitem{van2021active}
H.~Van~Gorp, I.~Huijben, B.~S. Veeling, N.~Pezzotti, and R.~J. Van~Sloun, ``Active deep probabilistic subsampling,'' in {\em International Conference on Machine Learning}, pp.~10509--10518, PMLR, 2021.

\bibitem{bahadir2019learning}
C.~D. Bahadir, A.~V. Dalca, and M.~R. Sabuncu, ``Learning-based optimization of the under-sampling pattern in {MRI},'' in {\em Information Processing in Medical Imaging: 26th International Conference, IPMI 2019, Hong Kong, China, June 2--7, 2019, Proceedings 26}, pp.~780--792, Springer, 2019.

\bibitem{zhang2020extending}
J.~Zhang, H.~Zhang, A.~Wang, Q.~Zhang, M.~Sabuncu, P.~Spincemaille, T.~D. Nguyen, and Y.~Wang, ``Extending {LOUPE} for k-space under-sampling pattern optimization in multi-coil {MRI},'' in {\em Machine Learning for Medical Image Reconstruction: Third International Workshop, MLMIR 2020, Held in Conjunction with MICCAI 2020, Lima, Peru, October 8, 2020, Proceedings 3}, pp.~91--101, Springer, 2020.

\bibitem{huijben2020learning}
I.~A. Huijben, B.~S. Veeling, and R.~J. van Sloun, ``Learning sampling and model-based signal recovery for compressed sensing {MRI},'' in {\em ICASSP 2020-2020 IEEE International Conference on Acoustics, Speech and Signal Processing (ICASSP)}, pp.~8906--8910, IEEE, 2020.

\bibitem{bengio2013estimating}
Y.~Bengio, N.~L{\'e}onard, and A.~Courville, ``Estimating or propagating gradients through stochastic neurons for conditional computation,'' {\em arXiv preprint arXiv:1308.3432}, 2013.

\bibitem{jang2016categorical}
E.~Jang, S.~Gu, and B.~Poole, ``Categorical reparameterization with gumbel-softmax,'' {\em arXiv preprint arXiv:1611.01144}, 2016.

\bibitem{aggarwal2020j}
H.~K. Aggarwal and M.~Jacob, ``{J-MoDL}: Joint model-based deep learning for optimized sampling and reconstruction,'' {\em IEEE journal of selected topics in signal processing}, vol.~14, no.~6, pp.~1151--1162, 2020.

\bibitem{weiss2019pilot}
T.~Weiss, O.~Senouf, S.~Vedula, O.~Michailovich, M.~Zibulevsky, and A.~Bronstein, ``{PILOT}: Physics-informed learned optimized trajectories for accelerated {MRI},'' {\em arXiv preprint arXiv:1909.05773}, 2019.

\bibitem{wang2022b}
G.~Wang, T.~Luo, J.-F. Nielsen, D.~C. Noll, and J.~A. Fessler, ``B-spline parameterized joint optimization of reconstruction and k-space trajectories ({BJORK}) for accelerated {2D} {MRI},'' {\em IEEE Transactions on Medical Imaging}, vol.~41, no.~9, pp.~2318--2330, 2022.

\bibitem{wang2022stochastic}
G.~Wang, J.-F. Nielsen, J.~A. Fessler, and D.~C. Noll, ``Stochastic optimization of {3D} non-cartesian sampling trajectory ({SNOPY}),'' {\em arXiv preprint arXiv:2209.11030}, 2022.

\bibitem{peng2022learning}
W.~Peng, L.~Feng, G.~Zhao, and F.~Liu, ``Learning optimal k-space acquisition and reconstruction using physics-informed neural networks,'' in {\em Proceedings of the IEEE/CVF Conference on Computer Vision and Pattern Recognition}, pp.~20794--20803, 2022.

\bibitem{beatty2005rapid}
P.~J. Beatty, D.~G. Nishimura, and J.~M. Pauly, ``Rapid gridding reconstruction with a minimal oversampling ratio,'' {\em IEEE transactions on medical imaging}, vol.~24, no.~6, pp.~799--808, 2005.

\bibitem{greengard2004accelerating}
L.~Greengard and J.-Y. Lee, ``Accelerating the nonuniform fast fourier transform,'' {\em SIAM review}, vol.~46, no.~3, pp.~443--454, 2004.

\bibitem{grover2019uncertainty}
A.~Grover and S.~Ermon, ``Uncertainty autoencoders: Learning compressed representations via variational information maximization,'' in {\em The 22nd International Conference on Artificial Intelligence and Statistics}, pp.~2514--2524, 2019.

\bibitem{linsker1988self}
R.~Linsker, ``Self-organization in a perceptual network,'' {\em Computer}, vol.~21, no.~3, pp.~105--117, 1988.

\bibitem{linsker1989generate}
R.~Linsker, ``How to generate ordered maps by maximizing the mutual information between input and output signals,'' {\em Neural computation}, vol.~1, no.~3, pp.~402--411, 1989.

\bibitem{barber2004algorithm}
D.~Barber and F.~Agakov, ``The im algorithm: a variational approach to information maximization,'' {\em Advances in neural information processing systems}, vol.~16, no.~320, p.~201, 2004.

\bibitem{kingma2013auto}
D.~P. Kingma and M.~Welling, ``Auto-encoding variational bayes,'' {\em arXiv preprint arXiv:1312.6114}, 2013.

\bibitem{mardani2018neural}
M.~Mardani, Q.~Sun, V.~Papyan, H.~Monajemi, S.~Vasanawala, J.~Pauly, and D.~Donoho, ``Neural proximal gradient descent for compressive imaging,'' {\em Advances in Neural Information Processing Systems}, vol.~31, pp.~9596--9606, 2018.

\bibitem{ong2018open}
F.~Ong, S.~Amin, S.~Vasanawala, and M.~Lustig, ``An open archive for sharing {MRI} raw data,'' in {\em ISMRM \& ESMRMB Joint Annu. Meeting}, p.~3425, 2018.

\bibitem{uecker2014espirit}
M.~Uecker, P.~Lai, M.~J. Murphy, P.~Virtue, M.~Elad, J.~M. Pauly, S.~S. Vasanawala, and M.~Lustig, ``{ESPIRiT}—an eigenvalue approach to autocalibrating parallel {MRI}: where {SENSE} meets {GRAPPA},'' {\em Magnetic resonance in medicine}, vol.~71, no.~3, pp.~990--1001, 2014.

\bibitem{ying2007joint}
L.~Ying and J.~Sheng, ``Joint image reconstruction and sensitivity estimation in {SENSE} ({JSENSE}),'' {\em Magnetic Resonance in Medicine: An Official Journal of the International Society for Magnetic Resonance in Medicine}, vol.~57, no.~6, pp.~1196--1202, 2007.

\bibitem{ong2019sigpy}
F.~Ong and M.~Lustig, ``{SigPy}: a python package for high performance iterative reconstruction,'' in {\em Proceedings of the ISMRM 27th Annual Meeting, Montreal, Quebec, Canada}, vol.~4819, p.~5, 2019.

\bibitem{baron2018rapid}
C.~A. Baron, N.~Dwork, J.~M. Pauly, and D.~G. Nishimura, ``Rapid compressed sensing reconstruction of 3d non-cartesian mri,'' {\em Magnetic resonance in medicine}, vol.~79, no.~5, pp.~2685--2692, 2018.

\bibitem{rasche1999resampling}
V.~Rasche, R.~Proksa, R.~Sinkus, P.~Bornert, and H.~Eggers, ``Resampling of data between arbitrary grids using convolution interpolation,'' {\em IEEE transactions on medical imaging}, vol.~18, no.~5, pp.~385--392, 1999.

\bibitem{wang2023efficient}
G.~Wang and J.~A. Fessler, ``Efficient approximation of jacobian matrices involving a non-uniform fast fourier transform ({NUFFT}),'' {\em IEEE Transactions on Computational Imaging}, vol.~9, pp.~43--54, 2023.

\bibitem{ramzi2021density}
Z.~Ramzi, J.-L. Starck, and P.~Ciuciu, ``Density compensated unrolled networks for non-{c}artesian {{MRI}} reconstruction,'' in {\em 2021 IEEE 18th International Symposium on Biomedical Imaging (ISBI)}, pp.~1443--1447, IEEE, 2021.

\bibitem{ramzi2022nc}
Z.~Ramzi, G.~Chaithya, J.-L. Starck, and P.~Ciuciu, ``{NC-PDNet}: A density-compensated unrolled network for {2D} and {3D} non-cartesian {MRI} reconstruction,'' {\em IEEE Transactions on Medical Imaging}, vol.~41, no.~7, pp.~1625--1638, 2022.

\bibitem{busse2008effects}
R.~F. Busse, A.~C. Brau, A.~Vu, C.~R. Michelich, E.~Bayram, R.~Kijowski, S.~B. Reeder, and H.~A. Rowley, ``Effects of refocusing flip angle modulation and view ordering in 3d fast spin echo,'' {\em Magnetic resonance in medicine}, vol.~60, no.~3, pp.~640--649, 2008.

\bibitem{sloane1984packing}
N.~J. Sloane, ``The packing of spheres,'' {\em Scientific American}, vol.~250, no.~1, pp.~116--125, 1984.

\bibitem{gossard2022spurious}
A.~Gossard, F.~de~Gournay, and P.~Weiss, ``Spurious minimizers in non uniform fourier sampling optimization,'' {\em Inverse Problems}, vol.~38, no.~10, p.~105003, 2022.

\bibitem{engel2021t}
M.~Engel, L.~Kasper, B.~Wilm, B.~Dietrich, L.~Vionnet, F.~Hennel, J.~Reber, and K.~P. Pruessmann, ``T-hex: Tilted hexagonal grids for rapid 3d imaging,'' {\em Magnetic Resonance in Medicine}, vol.~85, no.~5, pp.~2507--2523, 2021.

\bibitem{saranathan2007anthem}
M.~Saranathan, V.~Ramanan, R.~Gulati, and R.~Venkatesan, ``Anthem: anatomically tailored hexagonal mri,'' {\em Magnetic resonance imaging}, vol.~25, no.~7, pp.~1039--1047, 2007.

\bibitem{breuer2006controlled}
F.~A. Breuer, M.~Blaimer, M.~F. Mueller, N.~Seiberlich, R.~M. Heidemann, M.~A. Griswold, and P.~M. Jakob, ``Controlled aliasing in volumetric parallel imaging (2d caipirinha),'' {\em Magnetic Resonance in Medicine: An Official Journal of the International Society for Magnetic Resonance in Medicine}, vol.~55, no.~3, pp.~549--556, 2006.

\end{thebibliography}

\end{document}